\documentclass{article}
\usepackage{aip}

\input{tcilatex}

\begin{document}

\author{\textbf{H. I. Elim}$^{\thanks{%
hendry\_elim@hotmail.com}}$\textbf{\ \ \ and \ F. P. Zen}$^{\thanks{%
freddyzen@hotmail.com}}$ \\
$^{\ast }${\small Dept. of Physics, FMIPA, Pattimura Univ., Ambon,
Indonesia\ } \\
$^{\dagger }${\small Dept. of Theor. Physics, Bandung Institute of
Technology, Bandung, Indonesia}}
\title{Analysis of Hamiltonian Dynamics of Dispersion-Managed Coupled Breathers in
Optical Transmission System}
\date{9 June 2001 }
\maketitle

\begin{abstract}
We theoretically derive the Hamiltonian dynamics of dispersion-managed
coupled breathers in optical transmission system as a result \ of
phases-plane dynamics of the system. We analytically show that the
contribution of a perturbation term caused by a noise sourced by amplifiers
in optical fiber can influence the amplitudes and chirps of the pulses
dynamics. We also find that the coupled breathers dynamics depent on a
certain relationship of the chirps and amplitudes. The results of the
perturbed coupled\ \textbf{NLS} are then reduced\ to that in the unperturbed
system and the single \textbf{NLS} case.
\end{abstract}

\section{\textbf{Introduction}}

A dispersion-managed (\textbf{DM}) optical fiber is designed to create a low
(or even zero) path-averaged dispersion by periodically alternating
dispersion sign along an optical line which dramatically reduces pulse
broadening. Recently dispersion management has become an essential
technology for development of ultrafast high-bit rate optical communication
lines. There are some effective methods to improve the performance of a
soliton transmission system such as the use of a dispersion-compensation
technique, the use of dispersion-managed technique and so on.

Analytic techniques about the dispersion managed system of the single 
\textbf{NLS} by Kutz et. al.$^{1}$and Gabitov et al.$^{2}$ have been
separately developed to describe the amplitude and phase fluctuations that
occur on the scale of the dispersion map. The aim of the present paper is to
extend such analytic descriptions of leading-order amplitude and phase
fluctuations in the dispersion-managed coupled breathers derived from the
Gaussian pulses as a solution of the coupled NLS with a perturbation term of
a noise sourced by amplifier. We use an extended average variational method
to reduce the complicated coupled \textbf{NLS} with a periodic dispersion
map to several set of complicated nonlinear ordinary differential equations (%
\textbf{ODEs}), which accurately predict the completed coupled breathers
dynamics. In order to find the Hamiltonian dynamics of this system, the 
\textbf{ODEs} are derived by reducing and finding the parameters amplitudes $%
\eta _{j}\left( Z\right) ,$ and chirps $\beta _{j}\left( Z\right) $ in which
frequencies $f_{j}\left( Z\right) $ and center position of the pulses \c{T}$%
_{0_{j}}\left( Z\right) $ have been set to be identical with zero values.

The paper is organized as follows. In \textbf{Section 2} we introduce the
appropriate extended \textbf{coupled} \textbf{NLS with a perturbation term}
and its relevant physical parameters. In \textbf{Section 3} we derive the
Lagrangian of the nonintegrable coupled \textbf{NLS} with the perturbation
terms using variational method, and find that in order to make the method
work in this system, the parameters frequencies $f_{j}\left( Z\right) $ and
center position of the pulses \c{T}$_{0_{j}}\left( Z\right) $ must be chosen
to be very small ($\approx $ $0$).\textbf{\ Section 4} transforms the
extended coupled \textbf{NLS }equations from Lagrangian to Hamiltonian
formulation for which the solution of the equations of the coupled breathers
dynamics can be expressed in terms of a modified quadrature. \textbf{Section
5} contains of phase-plane dynamics of the derived coupled breathers
dynamical system. \textbf{Section 6} is the reduced results of section 5
related to the single \textbf{NLS} provided by Kutz, et al., and the
unperturbed coupled \textbf{NLS}. These system solutions are found by
reducing some parameters of the coupled \textbf{NLS} which has the
perturbation term.{\small \ }A conclusion of the results is provided in 
\textbf{Section 7}.

\section{\textbf{Formulation of the Nonintegrable Coupled NLS with a
Perturbation Term}}

A more complicated propagation of optical pulse in \textbf{DM} fiber is
described by a non-integrable coupled nonlinear Schr\"{o}dinger equation (%
\textbf{NLS}) with periodically varying dispersion\ $\sigma \left( Z\right) $
and some additional perturbation terms$:$\qquad 
\begin{equation}
\begin{array}{c}
i\frac{\partial Q_{1}}{\partial Z}+\frac{\sigma _{1}\left( Z\right) }{2}%
\frac{\partial ^{2}Q_{1}}{\partial T^{2}}+\left( \left| Q_{1}\right|
^{2}+\left| Q_{2}\right| ^{2}\right) Q_{1}=iG\left( Z\right) Q_{1}+i\epsilon
_{1}R_{1} \\ 
i\frac{\partial Q_{2}}{\partial Z}+\frac{\sigma _{2}\left( Z\right) }{2}%
\frac{\partial ^{2}Q_{2}}{\partial T^{2}}+\left( \left| Q_{1}\right|
^{2}+\left| Q_{2}\right| ^{2}\right) Q_{2}=iG\left( Z\right) Q_{2}+i\epsilon
_{2}R_{2}
\end{array}
\tag{1}
\end{equation}
where $Q_{1}\,$and $Q_{2}$ are the electric-field envelopes normalized by
the peak field powers $\left| E_{01}\right| ^{2}\,$\ and $\left|
E_{02}\right| ^{2}$ , $G\left( Z\right) =-\Gamma
+G_{a}\sum\limits_{n=1}^{M}\delta \left( Z-nZ_{a}\right) $ represents the
effects of the fiber loss and the amplification, where $\Gamma $ is the
normalized fiber loss, $G_{a}\left( Z\right) (\equiv \Gamma Z_{a})$ is the
gain of amplifier, the following term$^{3}$ 
\[
\begin{array}{c}
\epsilon _{j}R_{j}=r_{a_{j}}\exp \left\{ i\left[ f_{j}\left( Z\right) \left(
T-\text{\c{T}}_{0_{j}}\left( Z\right) \right) +\frac{\phi _{j}\left(
Z\right) }{2}\right] \right\} ,j=1,2
\end{array}
\]
are noise factors of the amplifiers ($1\gg \epsilon $) where $r_{a_{j}},$ $%
f_{j}\left( Z\right) ,$\c{T}$_{0_{j}}\left( Z\right) $ and $\phi _{j}\left(
Z\right) $ are complex amplitude of the noises, frequencies of the noises,
center position of the pulses, and the phases of the pulses, respectively
and we have set $f_{j}\left( Z\right) ,$and \c{T}$_{0_{j}}\left( Z\right) $
to be very small, and $\sigma _{j}\left( Z\right) $ is a periodic dispersion
map. To follow up, the previous work of Kutz, Holmes, $et$ $al.,$ in their
single \textbf{NLS} case, we define the variable $T$ as the physical time
normalized by $T_{0}/1.76$, where\ $T_{0}$ is the full width at half maximum
(\textbf{FWHM}) of the pulse ($T_{0}=50$ $ps$). The variable $Z$ represents
the physical distance devided by the dispersion length $Z_{0}$, which
corresponds to the average value $\left( \overline{D}\right) $ of the
dispersion map $\sigma \left( Z\right) $. Thus we find

\begin{equation}
Z_{0}=\frac{2\pi c}{\lambda _{0}\overline{D}}\left( \frac{T_{0}}{1.76}\right)
\tag{2}
\end{equation}
which gives

\begin{equation}
\left| E_{01}\right| ^{2}=\left| E_{02}\right| ^{2}\approx \frac{\lambda
_{0}A_{eff}}{2\pi n_{2}Z_{0}}  \tag{3}
\end{equation}
for the peak field intensities of the one-soliton solution corresponding to
the average dispersion. Here $n_{2}=2.6\times 10^{-16}$ cm$^{2}/$W is the
nonlinear coefficient of the fiber, $A_{eff}=55$ $\mu $m$^{2}$ is the
effective cross-sectional area of the fiber, and $\lambda _{0}=1.55$ $\mu $m
and $c$ are the carrier's free-space wavelength and speed of light,
respectively. In the present analysis, we involve the periodic intensity
fluctuations that are due to fiber loss and amplification because they have
very small perturbations to the average pulse dynamics. Thus the peak powers
given in eq.(3) are almost the same as the path-average peak power of the
propagating pulses. Note that we also consider long-length scale
fluctuations in intensity that are due to gain-shape changes and saturation
effects of the small noises in the amplifiers.

The parameters that are of greatest interest correspond to the particular
dispersion map that is being used. It is helpful therefore to consider the
specific case of a piecewise constant step-function dispersion map such that

\begin{equation}
\sigma _{1}\left( Z\right) =\sigma _{2}\left( Z\right) =\frac{1}{\overline{D}%
}{\Huge \{} 
\begin{array}{c}
D_{-},\,\ \ \ \ \ \ \ \ \ \ \ \ \ \ \ \ \ \ \ \ \ \ \ \ \ 0<Z<\frac{1}{2}%
\frac{Z_{-}}{Z_{0}} \\ 
D_{+},\text{ \ \ \ \ \ \ \ \ \ \ \ }\frac{1}{2}\frac{Z_{-}}{Z_{0}}<Z<\frac{1%
}{2}\frac{Z_{-}}{Z_{0}}+\frac{Z_{+}}{Z_{0}} \\ 
D_{-},\text{ \ \ }\frac{1}{2}\frac{Z_{-}}{Z_{0}}+\frac{Z_{+}}{Z_{0}}<Z<P=%
\frac{Z_{+}+Z_{-}}{Z_{0}}
\end{array}
\tag{4}
\end{equation}
where $\sigma \left( Z\right) =\sigma \left( Z+P\right) ,$ $D_{+}>0>D_{-}$
are the dispersions in each segment of fiber (in picoseconds per
kilometer.nanometer), $Z_{\pm }$ is the length of each fiber segments (in
kilometers), and the average dispersion is simply given by

\begin{equation}
\overline{D}=\frac{D_{+}Z_{+}+D_{-}Z_{-}}{Z_{-}+Z_{+}}.  \tag{5}
\end{equation}
We also follow the values of the parameters ($\overline{D}=0.2$ ps/km.nm, $%
Z_{-}=450$ km, $Z_{+}=60$ km, $D_{-}=-2.1$ ps/km.nm, and $D_{+}=17.45$
ps/km.nm) which are used by Kutz $et$ $\ al.$, in their explanation related
to the single \textbf{NLS}.

In order to simplify eq.(1), we use a transformation as follows

\begin{equation}
Q_{j}\left( Z,T\right) =\sqrt{\Upsilon (Z)}U_{j}\left( Z,T\right) ,\text{ \
\ \ }j=1,2.  \tag{6}
\end{equation}
We then find

\begin{equation}
\begin{array}{c}
i\frac{\partial U_{1}}{\partial Z}+\frac{\sigma \left( Z\right) }{2}\frac{%
\partial ^{2}U_{1}}{\partial T^{2}}+\Upsilon (Z)\left( \left| U_{1}\right|
^{2}+\left| U_{2}\right| ^{2}\right) U_{1}=\frac{i\epsilon _{1}R_{1}}{\sqrt{%
\Upsilon (Z)}} \\ 
i\frac{\partial U_{2}}{\partial Z}+\frac{\sigma \left( Z\right) }{2}\frac{%
\partial ^{2}U_{2}}{\partial T^{2}}+\Upsilon (Z)\left( \left| U_{1}\right|
^{2}+\left| U_{2}\right| ^{2}\right) U_{2}=\frac{i\epsilon _{2}R_{2}}{\sqrt{%
\Upsilon (Z)}}
\end{array}
\tag{7}
\end{equation}
where the reduced noise factor used in this investigation is 
\[
\epsilon _{j}R_{j}=\sqrt{\eta _{j}}r_{a_{j}}\exp \left\{ i\frac{\phi
_{j}\left( Z\right) }{2}\right\} , 
\]
in which the amplitude factors $\sqrt{\eta _{j}}$ used here are to reduce
the influence of the noise in the chirp dynamics and $\Upsilon (Z)$ is
chosen by following the work of Okamawari, $et$ $al.^{3}$ in the case of
their single \textbf{NLS} of eq.(7),

\begin{equation}
\Upsilon (Z)\equiv \Upsilon (0)\exp \left[ 2\int\limits_{0}^{Z}G(Z^{\prime
})dZ^{\prime }\right] ,  \tag{8}
\end{equation}
and the average values of\ $\Upsilon (Z)$ and $\sigma \left( Z\right) $ are
assumed to be unity in the normalized unit. The imaginary term in the
perturbation term can also be made as a real number by choosing, for
instance, the complex amplitude of the noises $r_{a_{j}}=\exp (-0.5i\phi
_{j}\left( Z\right) ).$

\section{\textbf{Variational Method of the Coupled NLS}}

The fundamental physical principle involved in this approach is derived from
Lagrangian methods or Euler-Lagrangian equation for classical mechanics
written by Goldstein, H.$^{4}$ This requires the dynamics to minimize a
certain energy integral closely related to in Hamilton's principle. We can
restate the extended coupled \textbf{NLS} eq.(7) in terms of its variational
form by defining the Lagrangian

\begin{equation}
L=\int\limits_{-\infty }^{\infty }\pounds \left( U_{j,}U_{j}^{\ast }\right)
dT,\text{ \ }j=1,2  \tag{9}
\end{equation}
where $\pounds \left( U_{j,}U_{j}^{\ast }\right) $ is the Lagrangian density
for the eq.(7) given by

\begin{eqnarray}
\pounds \left( U_{j,}U_{j}^{\ast }\right) &=&\,\sum\limits_{j=1}^{2}\left[
i\left( U_{j}\frac{\partial U_{j}^{\ast }}{\partial Z}-U_{j}^{\ast }\frac{%
\partial U_{j}}{\partial Z}\right) +\sigma \left( Z\right) \left| \frac{%
\partial U_{j}}{\partial T}\right| ^{2}-\Upsilon \left( Z\right) \left|
U_{j}\right| ^{4}\right]  \nonumber \\
&&-\sum\limits_{j=1}^{2}\frac{2i\epsilon _{j}R_{j}}{\sqrt{\Upsilon \left(
Z\right) }}\left( U_{j}-U_{j}^{\ast }\right) -\left( 2\Upsilon \left(
Z\right) \left| U_{1}\right| ^{2}\left| U_{2}\right| ^{2}\right) . 
\TCItag{10}
\end{eqnarray}
In order to reproduce eq.(7) and its complex conjugate, the Lagrangian
density must be determined by requiring the following equation

\begin{equation}
\frac{\delta \pounds \left( U_{j,}U_{j}^{\ast }\right) }{\delta U_{j}^{\ast }%
}=\frac{\delta \pounds \left( U_{j,}U_{j}^{\ast }\right) }{\delta U_{j}}=0. 
\tag{11}
\end{equation}

In the reduced model developed here, we assume that the fundamental dynamics
occurs owing to changes in amplitudes and quadratic phase chirps and they
depend on very small noises sourced by amplifiers. Further, we assume a
solution of eq.(7) that has an extended Gaussian shape as follows 
\begin{equation}
\begin{array}{c}
U_{j}\left( Z,T\right) =A_{j}b_{j}\sqrt{\eta _{j}}\exp \left\{ -\left(
\kappa _{j}\eta _{j}T\right) ^{2}+i\left[ \beta _{j}\kappa _{j}^{2}T^{2}+%
\frac{\phi _{j}\left( Z\right) }{2}\right] \right\} ,\ \ j=1,2
\end{array}
\tag{12}
\end{equation}
where $\eta _{j}\left( Z\right) $ and $\beta _{j}\left( Z\right) $ are free
parameters that correspond to the amplitude and quadratic-chirp dynamics,
respectively. We then find the total energy pulses as follows

\begin{eqnarray}
&&\int\limits_{-\infty }^{\infty }\left( \left| Q_{1}\left( Z,T\right)
\right| ^{2}+\left| Q_{2}\left( Z,T\right) \right| ^{2}\right) dT  \nonumber
\\
&=&\int \left[ \left( b_{1}^{2}\exp \left( -2\kappa _{1}^{2}T^{2}\right)
\right) +\left( b_{2}^{2}\exp \left( -2\kappa _{2}^{2}T^{2}\right) \right) %
\right] dT  \nonumber \\
&=&\frac{b_{1}^{2}}{\kappa _{1}}\sqrt{\frac{\pi }{2}}+\frac{b_{2}^{2}}{%
\kappa _{2}}\sqrt{\frac{\pi }{2}}  \TCItag{13}
\end{eqnarray}
As with the governing eq.(1), this ansatz conserves the total energy of the
pulse, i.e., $\int\limits_{-\infty }^{\infty }\left( \left| Q_{1}\right|
^{2}+\left| Q_{2}\right| ^{2}\right) dT=\left( \frac{A_{1}^{2}b_{1}^{2}}{%
\kappa _{1}}\sqrt{\frac{\pi }{2}}\right) +\left( \frac{A_{2}^{2}b_{2}^{2}}{%
\kappa _{2}}\sqrt{\frac{\pi }{2}}\right) $ = constant. More generally, any
localized ansatz for which the amplitude and width are related as in eq.(12)
can be utilized. Related to the case in the same width (\textbf{FWHM}) as
the one soliton solution, we find $\left( \frac{A_{1}^{2}b_{1}^{2}}{\kappa
_{1}}\sqrt{\frac{\pi }{2}}\right) +\left( \frac{A_{2}^{2}b_{2}^{2}}{\kappa
_{2}}\sqrt{\frac{\pi }{2}}\right) $ $=4$.

In general, approximating the eigenmode of eq.(1) with the dispersion map $%
\sigma \left( Z\right) $ by a Gaussian profile leads to errors in the
reduced dynamics. However these errors can be reduced by an appropriate
choice of the parameter $A_{j}$. To extend the previous work of the single 
\textbf{NLS} provided by Smith $et$ $al.^{5}$, we define the modified
amplitudes-enhancement factor $A_{j}$ as follows

\begin{equation}
A_{j}^{2}=1+\alpha _{j}\left\{ \frac{\lambda _{0}^{2}}{2\pi cT_{0}^{2}}\left[
\left( D_{+}-\overline{D}\right) Z_{+}-\left( D_{-}-\overline{D}\right) Z_{-}%
\right] \right\} ^{2},\text{ \ }j=1,2  \tag{14}
\end{equation}
The amplitudes-enhancement factor $\alpha _{j}\neq 0$, if we take an
arbitrary $A_{1}^{2}\neq A_{2}^{2}$ or $\left( A_{1}^{2}=A_{2}^{2}\right)
\neq 1$ in order to characterize a more complicated the Gaussian pulses
dynamics. This problem can also be easily discussed using a numerical
computation and simulation. However, in order to simplify our analytical
calculation we choose $A_{1}=A_{2}=A$ and then find $\kappa _{1}=\kappa _{2}=%
\frac{1.18}{1.76}$ if $b_{2}^{2}=b_{1}^{2}\approx 1.\,\allowbreak 070\,2.$

After substituting the pulses ansatz of eq.(12) in the Lagrangian equations
of eqs. (9) and (10) and evaluating the Lagrangian, we then find

\begin{eqnarray}
L &=&\sum\limits_{j=1}^{2}\frac{A_{j}^{2}b_{j}^{2}}{\kappa _{j}}\left\{ 
\frac{\sqrt{\frac{\pi }{2}}}{2\eta _{j}^{2}}\frac{\partial \beta _{j}}{%
\partial Z}+\left( \sqrt{\frac{\pi }{2}}\right) \frac{\partial \phi
_{j}\left( Z\right) }{\partial Z}-\frac{\Upsilon \left( Z\right) \sqrt{\pi }%
\eta _{j}}{2}\left( A_{j}^{2}b_{j}^{2}\right) \right\}  \nonumber \\
&&+\sum\limits_{j=1}^{2}\frac{A_{j}^{2}b_{j}^{2}}{\kappa _{j}}\left\{ \sqrt{%
\frac{\pi }{2}}\sigma _{j}\left( Z\right) \kappa _{j}^{2}\left[ \eta
_{j}^{2}+\left( \frac{\beta _{j}^{2}}{\eta _{j}^{2}}\right) \right] \right\}
\nonumber \\
&&-2\Upsilon \left( Z\right) \eta _{1}\eta
_{2}A_{1}^{2}A_{2}^{2}b_{1}^{2}b_{2}^{2}\sqrt{\frac{\pi }{2}}\left( \exp %
\left[ \left( \kappa _{1}^{2}\eta _{1}^{2}+\kappa _{2}^{2}\eta
_{2}^{2}\right) \right] \right)  \nonumber \\
&&-\left\{ \frac{i2\sqrt{\pi }}{\sqrt{\Upsilon \left( Z\right) }}\left(
r_{a_{j}}\Omega _{j}\right) \right\} ,  \TCItag{15}
\end{eqnarray}
where we have chosen

\[
\func{Re}\left( \left( \kappa _{j}\eta _{j}\right) ^{2}\right) >0 
\]
or

\[
\func{Re}\left( \left( \kappa _{j}\eta _{j}\right) ^{2}\right) =0\text{ \ \
\ \ \ and \ \ \ \ \ \ }\func{Im}\left( \left( \kappa _{j}\eta _{j}\right)
^{2}\right) >0, 
\]
and found

\begin{equation}
\Omega _{j}=\sum\limits_{j=1}^{2}\frac{A_{j}^{2}b_{j}^{2}}{\kappa _{j}}%
\left( \omega _{1_{j}}-\omega _{2_{j}}\right) ,\text{ \ }  \tag{16}
\end{equation}
where

\begin{equation}
\left( \omega _{1_{j}}-\omega _{2_{j}}\right) =\exp \left( \frac{-\beta
_{j}^{2}\kappa _{j}^{2}}{4\eta _{j}^{2}}\right) \left[ \exp \left( i\phi
_{j}\left( Z\right) \right) -1\right]  \tag{17}
\end{equation}

Applying the variational method with the Gaussian trial function of eq.(12),
we obtain that the related parameters of \c{T}$_{0_{j}}$ ,and $f_{i}$ in the
noise terms are very small and equal to zero.

Although the complicated variations with respect to the free parameters $%
\eta _{j}$, and $\beta _{j}$ can also yield the appropriate ordinary
differential equations (\textbf{ODEs}), we will extend the Lagrangian
formulation of the perturbative coupled \textbf{NLS }and will then try to
develop the Hamiltonian formulation of these equations by means of a Legendre%
\textbf{\ }transformation.This method reveals a conserved quantity, which is
utilized in the solution of the nonintegrable coupled \textbf{NLS} equation.
The validity of the variational method and its associated dynamics are also
in order related to the certain Gaussian pulse ansatz such as provided in
eq.(12). Then, we can produce the correct pulses evolution. This certainly
is not true in the case for which self-phases modulation dominates
dispersion and temporal side lobes are generated, i.e., if $A_{j}$ are too
large.

Although Kutz $et$ $al$. had already said that the variational method was 
\textbf{incapable} of describing or capturing the slight growth of
continuous spectra, which is generated by the dispersion-management
perturbations. We improve that it is possible if we can reduce all of the
additional parameters such as the frequencies $f_{j}\left( Z\right) $ and
center position of the pulses \c{T}$_{0_{j}}\left( Z\right) $ to be
identical with zero.

\section{The Hamiltonian Dynamics of The Coupled NLS}

There are various ways and manners in deriving the solution and explanation
of the amplitudes and chirps of the nonintegrable coupled \textbf{NLS}.
However for those cases in the nonintegrable systems, almost all of the
methods cannot work to derive the solutions. So, the only useful method we
are going to use is the Hamiltonian formulation of the dynamics. The
Hamiltonian system for the system is useful since solutions can be derived
and represented in terms of a modified quadrature. The Lagrangian system of
the preceding section is brought to Hamiltonian from via a Lengendre
transformation. We can directly define the related canonical coordinates $%
\beta _{j}$ with its conjugate variables $p_{\beta _{j}}$ as follows

\begin{equation}
p_{\beta _{j}}=\frac{\partial L}{\partial \left( \frac{d\beta _{j}}{dZ}%
\right) }=\frac{A_{j}^{2}}{\eta _{j}^{2}},\text{ \ \ \ \ }j=1,2  \tag{18}
\end{equation}
which represents the canonical momentums. The Legendre transformation
defines the Hamiltonian

\begin{equation}
H\left( \beta _{j},p_{\beta _{j}}\right) =p_{\beta _{j}}\frac{d\beta _{j}}{dZ%
}-L\left( \beta _{j},p_{\beta _{j}}\right) ,  \tag{19}
\end{equation}
where we must change the Lagrangian in eq.(15) into the Lagrangian in the
new canonical variables $\beta _{j}$ and $p_{\beta _{j}}$ and then
substitute it into eq.(19), we obtain

\begin{eqnarray}
H\left( \beta _{j},p_{\beta _{j}}\right) &=&\sum\limits_{j=1}^{2}\left(
p_{\beta _{j}}\frac{d\beta _{j}}{dZ}-p_{\beta _{j}}b_{j}^{2}\sqrt{\frac{\pi 
}{2}}\sigma _{j}\left( Z\right) \kappa _{j}\left[ \eta _{j}^{4}+\beta
_{j}^{2}\right] \right)  \nonumber \\
&&-\sum\limits_{j=1}^{2}\frac{p_{\beta _{j}}b_{j}^{2}\sqrt{\frac{\pi }{2}}}{%
\kappa _{j}}\left\{ \frac{1}{2}\frac{d\beta _{j}}{dZ}+\eta _{j}^{2}\frac{%
d\phi _{j}\left( Z\right) }{dZ}-\frac{\Upsilon \left( Z\right) \eta
_{j}^{5}b_{j}^{2}}{\sqrt{2}}\left( p_{\beta j}\right) \right\}  \nonumber \\
&&+2\sqrt{\frac{\pi }{2}}\Upsilon \left( Z\right) \left( p_{\beta
_{1}}\right) \left( p_{\beta _{2}}\right) \eta _{1}^{3}\eta
_{2}^{3}b_{1}^{2}b_{2}^{2}\exp \left( \kappa _{1}^{2}\eta _{1}^{2}+\kappa
_{2}^{2}\eta _{2}^{2}\right)  \nonumber \\
&&+\left( \frac{i2\sqrt{\pi }}{\sqrt{\Upsilon \left( Z\right) }}\right)
\left\{ r_{a_{1}}\Omega _{1}\left( \beta _{1},p_{\beta _{1}}\right)
+r_{a_{2}}\Omega _{2}\left( \beta _{2},p_{\beta _{2}}\right) \right\} , 
\TCItag{20}
\end{eqnarray}
where $\Omega _{j}\left( \beta _{j},p_{\beta _{j}}\right) $ are defined as
follows

\begin{equation}
\Omega _{j}\left( \beta _{j},p_{\beta _{j}}\right) =\frac{p_{\beta
_{j}}b_{j}^{2}\eta _{j}^{2}}{\kappa _{j}}\exp \left( \frac{-\beta
_{j}^{2}\kappa _{j}^{2}}{4\eta _{j}^{2}}\right) \left[ \exp \left( i\phi
_{j}\left( Z\right) \right) -1\right] .  \tag{21}
\end{equation}
The Hamiltonian can then be calculated with eq.(19), and the evolution
equations for $\eta _{j}$ dan $\beta _{j}$ are determined from the following
Hamilton's equations

\begin{equation}
\begin{array}{cc}
\frac{d\beta _{j}}{dZ}=\frac{\partial H\left( \beta _{j},p_{\beta
_{j}}\right) }{\partial p_{\beta _{j}}}, & \frac{dp_{\beta _{j}}}{dZ}=-\frac{%
\partial H\left( \beta _{j},p_{\beta j}\right) }{\partial \beta _{j}}.
\end{array}
\tag{22}
\end{equation}
In terms of $\eta _{j}$ , we can evaluate the evolution equations in eq.(22)
to be four \textbf{ODEs} stated the dynamical system of the Gaussian pulses
as follows

\begin{eqnarray}
\frac{\partial \beta _{1}}{\partial Z} &=&2\sigma _{1}\left( Z\right) \kappa
_{1}^{2}\left[ \eta _{1}^{4}-\beta _{1}^{2}\right] -\frac{1}{\sqrt{2}}%
\Upsilon \left( Z\right) \eta _{1}^{3}A_{1}^{2}b_{1}^{2}  \nonumber \\
&&-2\Upsilon \left( Z\right) \eta _{2}\eta _{1}^{3}A_{2}^{2}b_{2}^{2}\kappa
_{1}\left( 1+2\eta _{1}^{2}\kappa _{1}^{2}\right) \exp \left( \kappa
_{1}^{2}\eta _{1}^{2}+\kappa _{2}^{2}\eta _{2}^{2}\right)  \nonumber \\
&&-\left( \frac{\kappa _{1}}{A_{1}^{2}b_{1}^{2}}\right) \frac{i2\sqrt{2}%
r_{a_{1}}}{\sqrt{\Upsilon \left( Z\right) }}\left\{ \frac{\partial }{%
\partial \eta _{1}}\Omega _{1}\right\} ,  \TCItag{23a}
\end{eqnarray}

\begin{eqnarray}
\frac{\partial \beta _{2}}{\partial Z} &=&2\sigma _{2}\left( Z\right) \kappa
_{2}^{2}\left[ \eta _{2}^{4}-\beta _{2}^{2}\right] -\frac{\Upsilon \left(
Z\right) }{\sqrt{2}}\eta _{2}^{3}A_{2}^{2}b_{2}^{2}  \nonumber \\
&&-2\Upsilon \left( Z\right) \eta _{1}\eta _{2}^{3}A_{1}^{2}b_{1}^{2}\kappa
_{2}\left( 1+2\eta _{2}^{2}\kappa _{2}^{2}\right) \exp \left( \kappa
_{1}^{2}\eta _{1}^{2}+\kappa _{2}^{2}\eta _{2}^{2}\right)  \nonumber \\
&&-\left( \frac{\kappa _{2}}{A_{2}^{2}b_{2}^{2}}\right) \frac{i2\sqrt{2}%
r_{a_{2}}}{\sqrt{\Upsilon \left( Z\right) }}\left\{ \frac{\partial }{%
\partial \eta _{2}}\Omega _{2}\right\} ,  \TCItag{23b}
\end{eqnarray}
and

\begin{eqnarray}
\frac{d\eta _{j}}{dZ} &=&-b_{j}^{2}\sqrt{\frac{\pi }{2}}\sigma _{j}\left(
Z\right) \kappa _{j}\beta _{j}\eta _{j}  \nonumber \\
&&+\frac{\eta _{j}^{3}}{A_{j}^{2}}\frac{i\sqrt{\pi }r_{a_{j}}}{\sqrt{%
\Upsilon \left( Z\right) }}\left( \frac{\partial }{\partial \beta _{i}}%
\Omega _{j}\left( \beta _{j},p_{\beta _{j}}\right) \right) ,\text{ \ \ \ \ \
\ \ \ \ }j=1,2  \TCItag{23c}
\end{eqnarray}
where

\begin{equation}
\frac{\partial }{\partial \eta _{j}}\Omega _{j}=\frac{\beta _{j}^{2}\kappa
_{j}A_{j}^{2}b_{j}^{2}}{2\eta _{j}^{3}}\left( e^{\left( i\phi _{j}\left(
Z\right) \right) }-1\right) e^{^{\frac{-\left( \beta _{j}\kappa _{j}\right)
^{2}}{4\eta _{j}^{2}}}}  \tag{24a}
\end{equation}
and

\begin{equation}
\frac{\partial }{\partial \beta _{j}}\Omega _{j}=\frac{-\beta _{j}\kappa
_{j}A_{j}^{2}b_{j}^{2}}{2\eta _{j}^{2}}\left( e^{\left( i\phi _{j}\left(
Z\right) \right) }-1\right) e^{^{\frac{-\left( \beta _{j}\kappa _{j}\right)
^{2}}{4\eta _{j}^{2}}}}  \tag{24b}
\end{equation}

The results of the \textbf{ODEs} in eqs.(23) describe that the coupled
pulses dynamics depends on the chirps interactions contributed to the pulses
amplitudes. Loss and gain fluctuations related to the center position of the
pulses and the frequencies of noises are set very small to this formalism.
Hence, the system of equations decribes the nonlinear enhancement amplitudes
and width fluctuations along with the modified quadratic-chirp variations as
the pulses propagate\bigskip\ through a given dispersion map. Note that the
resulting \textbf{ODEs} system is identical to that derived by taking the
variations 
\begin{equation}
\frac{\delta L}{\delta \eta _{j}}=0\text{ \ \ \ \ \ \ and \ \ \ \ \ \ \ \ \
\ }\frac{\delta L}{\delta \beta _{j}}=0.  \tag{25}
\end{equation}

Since the resulting nonlinear \textbf{ODEs} system of eqs.(23) is the
Hamiltonian, and $\sigma _{j}\left( Z\right) $ is piecewise constant, we can
utilize the fact that

\begin{equation}
\frac{dH}{dZ}=0  \tag{26}
\end{equation}
and construct level sets of a scaled version of $H$ rewritten in terms of $%
\eta _{j}$ and $\beta _{j}$, 
\begin{eqnarray}
&& 
\begin{array}{c}
C=\left( \frac{\kappa _{1}}{b_{1}^{2}\sqrt{\frac{\pi }{2}}}-\frac{1}{2}%
\right) \frac{1}{\eta _{1}^{2}}\frac{d\beta _{1}}{dZ}+\left( \frac{\kappa
_{1}A_{2}^{2}}{A_{1}^{2}b_{1}^{2}\sqrt{\frac{\pi }{2}}}-\frac{A_{2}^{2}\frac{%
b_{2}^{2}}{\kappa _{2}}}{2A_{1}^{2}\frac{b_{1}^{2}}{\kappa _{1}}}\right) 
\frac{1}{\eta _{2}^{2}}\frac{d\beta _{2}}{dZ} \\ 
-\left\{ \frac{d\phi _{1}\left( Z\right) }{dZ}+\frac{A_{2}^{2}\frac{b_{2}^{2}%
}{\kappa _{2}}}{A_{1}^{2}\frac{b_{1}^{2}}{\kappa _{1}}}\frac{d\phi
_{2}\left( Z\right) }{dZ}\right\} -\sigma _{1}\left( Z\right) \kappa _{1}^{2}%
\left[ \eta _{1}^{2}+\frac{\beta _{1}^{2}}{\eta _{1}^{2}}\right] \\ 
-\frac{A_{2}^{2}b_{2}^{2}\kappa _{1}}{A_{1}^{2}b_{1}^{2}}\sigma _{2}\left(
Z\right) \kappa _{2}\left[ \eta _{2}^{2}+\frac{\beta _{2}^{2}}{\eta _{2}^{2}}%
\right] \\ 
+\left( \frac{\Upsilon \left( Z\right) }{\sqrt{2}}\right) \left( \eta
_{1}A_{1}^{2}b_{1}^{2}\right) +\frac{A_{2}^{4}\frac{b_{2}^{2}}{\kappa _{2}}}{%
A_{1}^{2}\frac{b_{1}^{2}}{\kappa _{1}}}\left( \frac{\Upsilon \left( Z\right) 
}{\sqrt{2}}\right) \left( \eta _{2}b_{2}^{2}\right) \\ 
+2\Upsilon \left( Z\right) A_{2}^{2}\kappa _{1}\eta _{1}\eta
_{2}b_{2}^{2}\left( \exp \left[ \left( \kappa _{1}^{2}\eta _{1}^{2}+\kappa
_{2}^{2}\eta _{2}^{2}\right) \right] \right) \\ 
+\frac{1}{A_{1}^{2}b_{1}^{2}}\left( \frac{i2\sqrt{2}\kappa _{1}}{\sqrt{%
\Upsilon \left( Z\right) }}\right) \left\{ r_{a_{1}}\Omega _{1}\left( \beta
_{1},p_{\beta _{1}}\right) +r_{a_{2}}\Omega _{2}\left( \beta _{1},p_{\beta
_{1}}\right) \right\} ,
\end{array}
\nonumber \\
&&  \TCItag{27}
\end{eqnarray}
In order to reduce the eq.(27), we choose 
\begin{equation}
b_{2}=\left( \frac{2\kappa _{2}}{\sqrt{\frac{\pi }{2}}}\right) ^{\frac{1}{2}}%
\text{, \ \ \ \ \ \ \ and \ }b_{1}=\left( \frac{\left( 4-2A_{2}^{2}\right)
\kappa _{1}}{A_{1}^{2}\sqrt{\frac{\pi }{2}}}\right) ^{\frac{1}{2}}\text{\ .}
\tag{28a}
\end{equation}
The eq.(27) can be more simplified by choosing

\begin{equation}
b_{j}^{2}=2\sqrt{\frac{2}{\pi }}\kappa _{j}=b^{2}\approx 1.\,\allowbreak
070\,2,  \tag{28b}
\end{equation}
and $\left( \kappa _{1}=\kappa _{2}=\kappa =\frac{1.18}{1.76}\right) $

\begin{equation}
A_{1}^{2}=A_{2}^{2}=1.  \tag{28c}
\end{equation}
We then find the general solution of $\beta _{1}\left( \eta _{1},\eta
_{2},\beta _{2}\right) $ as follows

\begin{equation}
\begin{array}{c}
\beta _{1}=\pm \eta _{1}\left\{ \gamma _{1}\left[ \eta _{1}+\eta _{2}\right]
-\gamma _{2}-\gamma _{3}\left( \eta _{2}^{2}+\frac{\beta _{2}^{2}}{\eta
_{2}^{2}}\right) +\gamma _{4}\eta _{1}\eta _{2}e^{\left[ \kappa ^{2}\left(
\eta _{1}^{2}+\eta _{2}^{2}\right) \right] }+\gamma _{5}-\eta
_{1}^{2}\right\} ^{\frac{1}{2}},
\end{array}
\tag{29}
\end{equation}
where

\begin{eqnarray}
\gamma _{1} &=&\left( \frac{2\Upsilon \left( Z\right) }{\sigma \left(
Z\right) \kappa \sqrt{\pi }}\right) ,  \TCItag{30a} \\
\gamma _{2} &=&\left( \frac{C+\frac{d\phi _{1}\left( Z\right) }{dZ}+\frac{%
d\phi _{2}\left( Z\right) }{dZ}}{\sigma _{1}\left( Z\right) \kappa ^{2}}%
\right) ,  \TCItag{30b} \\
\gamma _{3} &=&1\text{ \ \ \ \ \ \ \ }\left( \text{for \ }\sigma _{1}\left(
Z\right) =\sigma _{2}\left( Z\right) =\sigma \right) ,  \TCItag{30c} \\
\gamma _{4} &=&\frac{4\sqrt{2}\Upsilon \left( Z\right) }{\sigma \left(
Z\right) \sqrt{\pi }},  \TCItag{30d}
\end{eqnarray}
and

\begin{equation}
\gamma _{5}=\left( \frac{i\sqrt{\pi }}{\sigma \left( Z\right) \kappa ^{2}%
\sqrt{\Upsilon \left( Z\right) }}\right) \left[ r_{a_{1}}\Omega _{1}\left(
\beta _{1},p_{\beta _{1}}\right) +r_{a_{2}}\Omega _{2}\left( \beta
_{2},p_{\beta _{2}}\right) \right]  \tag{30e}
\end{equation}
where $\Omega _{1}\left( \beta _{1},p_{\beta _{1}}\right) $ \ and $\Omega
_{2}\left( \beta _{2},p_{\beta _{2}}\right) $ are defined in eq.(21).

\section{Phases-Plane Dynamics of The Perturbed Coupled NLS}

It is often convenient and insightful to plot the phase-plane dynamics of a
planar dynamical system such as eqs.(23). This geometric representation of
the solution treats the distance $Z$ as a parameter while plotting the
amplitudes $\eta _{j}$ and chirps $\beta _{j}$ variables on the $x$ and $y$
axes, respectively. It elucidates some fundamental aspects of the coupled
breathers dynamics, which will be clearly shown in what follows.

The phases-plane dynamics are markedly different depending on the sign of
the dispersion. Thus we consider separately the anomalous- and
normal-dispersion regimes. Before seeking the fixed points of eqs.(23), we
will simplify the very complicated equations in order to get the
phases-plane dynamics consisted of $\beta _{1}\left( \eta _{1}\right) $, $%
\beta _{1}\left( \eta _{2}\right) $, $\beta _{2}\left( \eta _{1}\right) $
and $\beta _{2}\left( \eta _{1}\right) .$ This will easily work by choosing

\begin{equation}
\beta _{2}=\rho _{1}\left( Z\right) \beta _{1},  \tag{31a}
\end{equation}
and

\begin{equation}
\eta _{2}=\rho _{2}\left( Z\right) \eta _{1},  \tag{31b}
\end{equation}
or by setting $\rho _{1}\left( Z\right) =\rho _{2}\left( Z\right) $, we
suggest and make our solution in the simplest forms: 
\begin{equation}
\rho \left( Z\right) =\frac{\beta _{2}}{\beta _{1}}=\frac{\eta _{2}}{\eta
_{1}}=const.\text{ \ \ \ \ }  \tag{31c}
\end{equation}
After substituting eq.(31c) into eqs.(23) and eq.(29), we then find

\begin{equation}
\beta _{1}=\pm \eta _{1}\left\{ \gamma _{a}\eta _{1}-\frac{\gamma _{2}}{\tau 
}-\gamma _{b}\eta _{1}^{2}+\gamma _{c}\eta _{1}^{2}\exp \left[ \kappa
^{2}\eta _{1}^{2}\left( 1+\rho ^{2}\right) \right] +\frac{\gamma _{5}}{\tau }%
-\frac{\eta _{1}^{2}}{\tau }\right\} ^{\frac{1}{2}},  \tag{32a}
\end{equation}

\begin{eqnarray}
\beta _{1} &=&\pm \eta _{2}\left\{ \frac{\gamma _{a}}{\rho ^{3}}\eta _{2}-%
\frac{\gamma _{2}}{\rho ^{2}\tau }-\frac{\gamma _{b}}{\rho ^{4}}\eta
_{2}^{2}+\frac{\gamma _{c}}{\rho ^{4}}\eta _{2}^{2}\left( e^{\kappa
^{2}\left( 1+\frac{1}{\rho ^{2}}\right) \eta _{2}^{2}}\right) +\frac{\gamma
_{5}}{\rho ^{2}\tau }-\frac{1}{\rho ^{4}\tau }\eta _{2}^{2}\right\} ^{\frac{1%
}{2}},  \nonumber \\
&&  \TCItag{32b}
\end{eqnarray}

\begin{eqnarray}
\beta _{2} &=&\pm \eta _{2}\left\{ \frac{\gamma _{a}\eta _{2}}{\rho }-\frac{%
\gamma _{2}}{\tau }-\frac{\gamma _{b}\eta _{2}^{2}}{\rho ^{2}}+\frac{\gamma
_{c}\eta _{2}^{2}\left( e^{\kappa ^{2}\left( 1+\frac{1}{\rho ^{2}}\right)
\eta _{2}^{2}}\right) }{\rho ^{2}}+\frac{\gamma _{5}}{\tau }-\frac{\eta
_{2}^{2}}{\rho ^{2}\tau }\right\} ^{\frac{1}{2}},  \nonumber \\
&&  \TCItag{32c}
\end{eqnarray}

\begin{eqnarray}
\beta _{2} &=&\pm \eta _{1}\left\{ \rho ^{2}\gamma _{a}\eta _{1}-\frac{\rho
^{2}\gamma _{2}}{\tau }-\rho ^{2}\gamma _{b}\eta _{1}^{2}+\rho ^{2}\gamma
_{c}\eta _{1}^{2}e^{\kappa ^{2}\eta _{1}^{2}\left( 1+\rho ^{2}\right) }+%
\frac{\rho ^{2}\gamma _{5}}{\tau }-\frac{\rho ^{2}\eta _{1}^{2}}{\tau }%
\right\} ^{\frac{1}{2}},  \nonumber \\
&&  \TCItag{32d}
\end{eqnarray}
where we have reduced the eq.(32) by choosing $\tau $ as follows $\left(
\sigma _{1}\left( Z\right) =\sigma _{2}\left( Z\right) \right) $

\begin{equation}
\tau =2,  \tag{33}
\end{equation}
and the parameters $\gamma _{a},$ $\gamma _{b},\gamma _{c},$and the revised $%
\gamma _{5}$ are

\begin{eqnarray}
\gamma _{a} &=&\frac{\gamma _{1}\left[ 1+\rho \left( Z\right) \right] }{2}, 
\TCItag{34a} \\
\gamma _{b} &=&\frac{\gamma _{3}\rho ^{2}\left( Z\right) }{2},  \TCItag{34b}
\\
\gamma _{c} &=&\frac{\gamma _{4}\rho \left( Z\right) }{2},  \TCItag{34c}
\end{eqnarray}
and

\begin{eqnarray}
\gamma _{5} &=&\left( \frac{i\sqrt{\pi }r_{a}}{\sigma \left( Z\right) \kappa
^{2}\sqrt{\Upsilon \left( Z\right) }}\right) \left[ \Omega _{1}+\Omega _{2}%
\right]  \nonumber \\
&=&\frac{i\sqrt{\pi }r_{a}b^{2}\Xi \left( Z\right) }{\sigma \left( Z\right)
\kappa ^{3}\sqrt{\Upsilon \left( Z\right) }}\left[ e^{\left( i\phi
_{1}\left( Z\right) +\right) }+e^{\left( i\phi _{2}\left( Z\right) \right)
}-2\right] ,\text{ for }r_{a_{1}}=r_{a_{2}}=r_{a},\text{ }  \nonumber \\
&&  \TCItag{34d}
\end{eqnarray}
or by choosing $r_{a_{j}}=\exp \left[ -0.5i\phi _{j}\left( Z\right) \right] $
in the eq.(34d), we can reduce the imaginary and then find 
\begin{equation}
\gamma _{5}=\left( \frac{-2\sqrt{\pi }b^{2}\Xi }{\sigma \kappa ^{3}\sqrt{%
\Upsilon }}\right) \left[ \sin \left( 0.5\phi _{1}\right) +\sin \left(
0.5\phi _{2}\right) \right]  \tag{34e}
\end{equation}
$\allowbreak $ $\allowbreak $where $\left( \text{\ }\Xi \left( Z\right)
>1\right) .$ In eq.(34d), we have suggested the following relationships

\begin{equation}
\Xi =-\frac{i8\sqrt{\Upsilon }\sigma \kappa }{\sqrt{\pi }r_{a}b^{2}\left(
e^{i\phi _{1}}+e^{i\phi _{2}}-2\right) }\func{LambertW}\left( \frac{1}{8}i%
\sqrt{\pi }r_{a}\frac{b^{2}}{\sqrt{\Upsilon }}\frac{e^{i\phi _{1}}+e^{i\phi
_{2}}-2}{\sigma \kappa }\left( \text{\c{L}}\right) \right) ,  \tag{35a}
\end{equation}
where

\[
\text{\c{L}}=\exp \left( -\frac{1}{4}\kappa ^{2}\gamma _{a}\eta _{1}-\frac{1%
}{4}\kappa ^{2}\gamma _{c}\eta _{1}^{2}e^{\left( \kappa ^{2}\eta
_{1}^{2}\left( 1+\rho ^{2}\right) \right) }+\frac{1}{8}\kappa ^{2}\gamma
_{2}+\frac{1}{4}\kappa ^{2}\eta _{1}^{2}\gamma _{b}+\frac{1}{8}\kappa
^{2}\eta _{1}^{2}\right) , 
\]
and

\begin{equation}
-\left( \frac{\kappa ^{2}}{4\eta _{1}^{2}}\right) \beta _{1}^{2}+\left( 
\frac{\kappa ^{2}}{8\eta _{1}^{2}}\right) \beta _{1}^{4}+O\left( \beta
_{1}^{6}\right) >0,  \tag{35b}
\end{equation}
and the rule appeared in eqs.(35a) and (35b) must be used to our entire
solutions in this section. If we choose $\Xi \left( Z\right) =1$ or the
chirp $\beta _{1}\approx 0$ , this then means that the pulse width is
minimum at the center of unit cell. However, the mean-square frequency shift
is not minimum. We can also suppose that a noise $\left( \text{in }\gamma
_{5}\text{ term}\right) $ sourced by the amplifier which has a small
frequency causes a small change in the coupled breathers dynamics of the
Gaussian pulses. However, we still can reduce the noise by enhancing the
amplitude of the pulses.

The aim of the use of a relationship in eq.(31c) is to make the phases-plane
dynamics of the coupled breathers are in stable conditions. A first order 
\textbf{ODE} in terms of $\eta _{1}$alone can then be derived by
substituting eq. (32a) into eq.(23c):

\begin{eqnarray}
\frac{d\eta _{1}}{dZ} &=&\mp \left( b^{2}\sqrt{\frac{\pi }{2}}\sigma
_{1}\left( Z\right) \kappa \eta _{1}^{2}\smallskip \right)  \nonumber \\
&&\times \left\{ \gamma _{a}\eta _{1}-\frac{\gamma _{2}}{\tau }-\gamma
_{b}\eta _{1}^{2}+\gamma _{c}\eta _{1}^{2}\exp \left[ \kappa ^{2}\eta
_{1}^{2}\left( 1+\rho _{2}^{2}\right) \right] +\frac{\gamma _{5}}{\tau }-%
\frac{\eta _{1}^{2}}{\tau }\right\} ^{\frac{1}{2}}  \nonumber \\
&&\mp \left( \frac{ir_{a_{1}}\sqrt{\pi }\Xi \left( Z\right) \left( e^{\left(
i\phi _{1}\left( Z\right) \right) }-1\right) \kappa b^{2}\eta _{1}^{2}}{2%
\sqrt{\Upsilon \left( Z\right) }}\right)  \nonumber \\
&&\times \left\{ \gamma _{a}\eta _{1}-\frac{\gamma _{2}}{\tau }-\gamma
_{b}\eta _{1}^{2}+\gamma _{c}\eta _{1}^{2}\exp \left[ \kappa ^{2}\eta
_{1}^{2}\left( 1+\rho _{2}^{2}\right) \right] +\frac{\gamma _{5}}{\tau }-%
\frac{\eta _{1}^{2}}{\tau }\right\} ^{\frac{1}{2}}.  \TCItag{36}
\end{eqnarray}
Upon integrating this equation with initial conditions $\eta _{1}=\eta
_{0_{1}}$ at $Z=0$, we find

\begin{eqnarray}
Z &=&\int\limits_{\eta _{01}}^{\eta _{1}}\frac{\mp \left\{ \gamma _{a}\eta
_{1}-\frac{\gamma _{2}}{\tau }-\gamma _{b}\eta _{1}^{2}+\gamma _{c}\eta
_{1}^{2}\exp \left[ \kappa ^{2}\eta _{1}^{2}\left( 1+\rho _{2}^{2}\right) %
\right] +\frac{\gamma _{5}}{\tau }-\frac{\eta _{1}^{2}}{\tau }\right\} ^{-%
\frac{1}{2}}d\eta _{1}}{\left\{ b^{2}\sqrt{\frac{\pi }{2}}\sigma _{1}\left(
Z\right) \kappa \smallskip +\frac{i\sqrt{\pi }\Xi \left( Z\right) \left(
e^{\left( i\phi _{1}\left( Z\right) \right) }-1\right) \kappa b^{2}r_{a_{1}}%
}{2\sqrt{\Upsilon \left( Z\right) }}\right\} \eta _{1}^{2}}  \nonumber \\
&&  \TCItag{37}
\end{eqnarray}
Another first order \textbf{ODE} in terms of $\eta _{2}$ alone will easily
be derived using the same manner. Although we already get an analytic
solution for the pulse dynamics in eq.(37), the integration is still very
complicated due to some closed-form expressions. The simplest way left to
explain the coupled breathers dynamics is a geometrical interpretation of
the \textbf{ODEs} in eqs.(23).

Finally, we come to obtain the fixed points of eqs.(23) by setting the
right-hand sides equal to zero. There are three fixed points which can be
proved by figuring the phases-plane dynamics consisted of $\beta _{1}\left(
\eta _{1}\right) $, $\beta _{1}\left( \eta _{2}\right) $, $\beta _{2}\left(
\eta _{1}\right) $ and $\beta _{2}\left( \eta _{1}\right) $ :

\begin{equation}
\mathbf{I.}\text{ \ \ \ }\beta _{j}=0,\text{ \ \ \ \ \ }\eta _{j}=0, 
\tag{38a}
\end{equation}

\begin{eqnarray}
\mathbf{II}.\text{ \ \ \ }\beta _{j} &=&0,\text{ \ \ \ \ and}  \nonumber \\
\eta _{1} &=&\frac{a_{21}a_{71}}{a_{71}a_{11}-a_{41}\exp \left( \vartheta
_{1}\right) a_{31}\vartheta _{1}-a_{71}\exp \left( \vartheta _{1}\right)
a_{31}}=\eta _{10},  \TCItag{38b}
\end{eqnarray}

\begin{eqnarray}
\mathbf{III.}\text{ \ \ \ }\beta _{j} &=&0,\text{\ \ \ \ and\ \ }  \nonumber
\\
\eta _{2} &=&\frac{c_{21}c_{71}}{c_{71}c_{11}-c_{41}\exp \left( \vartheta
_{1}^{\prime }\right) c_{31}\vartheta _{1}^{\prime }-c_{71}\exp \left(
\vartheta _{1}^{\prime }\right) c_{31}}=\eta _{20},  \TCItag{38c}
\end{eqnarray}
\bigskip where $\vartheta _{1}$ is a root of

\begin{eqnarray}
&&{\huge \{}e^{2\Theta }\Theta a_{71}^{2}a_{31}^{2}+2e^{2\Theta
}a_{71}a_{31}^{2}a_{41}\Theta ^{2}+e^{2\Theta }a_{41}^{2}a_{31}^{2}\Theta
^{3}-2e^{\Theta }\Theta a_{71}^{2}a_{11}a_{31}  \nonumber \\
&&-2e^{\Theta }a_{71}a_{11}a_{41}a_{31}\Theta
^{2}-a_{21}^{2}a_{71}^{3}+\Theta a_{71}^{2}a_{11}^{2}{\huge \}}, 
\TCItag{39a}
\end{eqnarray}
and $\vartheta _{1}^{\prime }$ is a root of 
\begin{eqnarray}
&&{\huge \{}e^{2\Theta }\Theta c_{71}^{2}c_{31}^{2}+2e^{2\Theta
}c_{71}c_{31}^{2}c_{41}\Theta ^{2}+e^{2\Theta }c_{41}^{2}c_{31}^{2}\Theta
^{3}-2e^{\Theta }\Theta c_{71}^{2}c_{11}c_{31}  \nonumber \\
&&-2e^{\Theta }c_{71}c_{11}c_{41}c_{31}\Theta
^{2}-c_{21}^{2}c_{71}^{3}+\Theta c_{71}^{2}c_{11}^{2}{\huge \}}, 
\TCItag{39b}
\end{eqnarray}
and the parameters $\eta _{01},\eta
_{02},a_{11},a_{21},a_{31},a_{41},a_{71}, $ $c_{11},c_{21},c_{31},c_{41},$%
and $c_{71}$ (for $A_{1}=A_{2}=A$ and $\sigma _{1}\left( Z\right) =\sigma
_{2}\left( Z\right) =\sigma \left( Z\right) $) are

\[
\begin{array}{c}
\eta _{01}=\frac{\frac{\Upsilon \left( Z\right) }{\sqrt{2}}A^{2}b^{2}\kappa
^{2}\left( 1+\rho ^{2}\right) }{2\kappa ^{4}\left( 1+\rho ^{2}\right) \sigma
-2\kappa ^{2}\exp \left( \vartheta _{1}\right) \left[ 2\Upsilon \left(
Z\right) \rho A^{2}b^{2}\kappa \right] \vartheta _{1}-\kappa ^{2}\left(
1+\rho ^{2}\right) \exp \left( \vartheta _{1}\right) \left[ 2\Upsilon \left(
Z\right) \rho A^{2}b^{2}\kappa \right] }, \\ 
\\ 
\eta _{02}=\frac{\frac{\Upsilon \left( Z\right) }{\sqrt{2}}A^{2}b^{2}\kappa
^{2}\left( 1+\frac{1}{\rho ^{2}}\right) }{2\kappa ^{4}\left( 1+\frac{1}{\rho
^{2}}\right) \sigma -2\kappa ^{2}\exp \left( \vartheta _{1}^{\prime }\right) 
\left[ 2\Upsilon \left( Z\right) \rho A^{2}b^{2}\kappa \right] \vartheta
_{1}^{\prime }-\kappa ^{2}\left( 1+\frac{1}{\rho ^{2}}\right) \exp \left(
\vartheta _{1}^{\prime }\right) \left[ 2\Upsilon \left( Z\right) \rho
A^{2}b^{2}\kappa \right] },
\end{array}
\]

\begin{eqnarray*}
a_{11} &=&c_{11}=2\sigma \left( Z\right) \kappa ^{2}, \\
a_{21} &=&c_{21}=\frac{A^{2}\Upsilon \left( Z\right) b^{2}}{\sqrt{2}}, \\
a_{31} &=&c_{31}=2\Upsilon \left( Z\right) \rho A^{2}b^{2}\kappa , \\
a_{41} &=&c_{41}=2\kappa ^{2}, \\
a_{71} &=&\kappa ^{2}\left( 1+\rho ^{2}\right) , \\
c_{71} &=&\kappa ^{2}\left( 1+\frac{1}{\rho ^{2}}\right) .
\end{eqnarray*}

A linear-stability calculation about each of the fixed points gives insight
into the general coupled breathers dynamics of the system. In particular,
the fixed points $\left( \eta _{j},\beta _{j}\right) =\left( 0,0\right) $\
are degenerate so that their linear stabilities cannot be determined (see,
for examples, figures 1, 3, 5, and 7). However, the linear stability of the
fixed point II and III can be determined by studying the pulses-plane
dymanics of the \textbf{ODEs} in eq.(23).\textbf{\ }Linearizing about these
points by setting

\begin{equation}
\eta _{j}=\eta _{0j}+\widetilde{\eta }_{j}  \tag{40a}
\end{equation}

\begin{equation}
\beta _{j}=0+\widetilde{\beta _{j}}  \tag{40b}
\end{equation}
yields the linear system and it can be rewritten in a matrix form as follows

\begin{equation}
\left( 
\begin{array}{c}
\frac{d\widetilde{\eta _{1}}}{dZ} \\ 
\frac{d\widetilde{\eta _{2}}}{dZ} \\ 
\frac{\partial \widetilde{\beta _{1}}}{\partial Z} \\ 
\frac{\partial \widetilde{\beta _{2}}}{\partial Z}
\end{array}
\right) =\left( 
\begin{array}{cccc}
0 & 0 & M_{13} & 0 \\ 
0 & 0 & 0 & M_{24} \\ 
M_{31} & 0 & 0 & 0 \\ 
0 & M_{42} & 0 & 0
\end{array}
\right) \left( 
\begin{array}{c}
\widetilde{\eta _{1}} \\ 
\widetilde{\eta _{2}} \\ 
\widetilde{\beta _{1}} \\ 
\widetilde{\beta _{2}}
\end{array}
\right) ,  \tag{41}
\end{equation}
where 
\begin{eqnarray}
M_{13} &=&-b^{2}\sqrt{\frac{\pi }{2}}\sigma _{1}\left( Z\right) \kappa \eta
_{01}  \nonumber \\
&&-\frac{ir_{a_{1}}\sqrt{\pi }\left( e^{\left( i\phi _{1}\left( Z\right)
\right) }-1\right) \kappa b^{2}\Xi \left( Z\right) \eta _{01}}{2\sqrt{%
\Upsilon \left( Z\right) }}  \TCItag{42a}
\end{eqnarray}

\begin{eqnarray}
M_{24} &=&-b^{2}\sqrt{\frac{\pi }{2}}\sigma _{2}\left( Z\right) \kappa \eta
_{02}  \nonumber \\
&&-\frac{ir_{a_{2}}\sqrt{\pi }\left( e^{\left( i\phi _{2}\left( Z\right)
\right) }-1\right) \kappa b^{2}\Xi \left( Z\right) \eta _{02}}{2\sqrt{%
\Upsilon \left( Z\right) }}  \TCItag{42b}
\end{eqnarray}

\begin{eqnarray}
M_{31} &=&8\sigma _{1}\left( Z\right) \kappa ^{2}\eta _{01}^{3}-\frac{3}{%
\sqrt{2}}\Upsilon \left( Z\right) A_{1}^{2}b^{2}\eta _{01}^{2}  \nonumber \\
&&-8\Upsilon \left( Z\right) \rho A_{2}^{2}b^{2}\kappa \eta _{01}^{3}\left(
1+2\eta _{01}^{2}\kappa ^{2}\right) e^{\left( \kappa ^{2}\left( 1+\rho
^{2}\right) \eta _{01}^{2}\right) }  \nonumber \\
&&-8\Upsilon \left( Z\right) \rho A_{2}^{2}b^{2}\kappa ^{3}\eta
_{01}^{5}e^{\left( \kappa ^{2}\left( 1+\rho ^{2}\right) \eta
_{01}^{2}\right) }  \nonumber \\
&&-4\Upsilon \left( Z\right) \rho A_{2}^{2}b^{2}\kappa ^{3}\eta
_{01}^{5}\left( 1+2\eta _{01}^{2}\kappa ^{2}\right) \left( 1+\rho
^{2}\right) e^{\left( \kappa ^{2}\left( 1+\rho ^{2}\right) \eta
_{01}^{2}\right) }  \TCItag{42c}
\end{eqnarray}

\begin{eqnarray}
&&  \nonumber \\
&&  \nonumber \\
&&  \nonumber \\
M_{42} &=&8\sigma _{2}\left( Z\right) \kappa ^{2}\eta _{02}^{3}-\frac{%
3\Upsilon \left( Z\right) }{\sqrt{2}}A_{2}^{2}b^{2}\eta _{02}^{2}  \nonumber
\\
&&-\frac{8\Upsilon \left( Z\right) }{\rho }A_{1}^{2}b^{2}\kappa \eta
_{02}^{3}\left( 1+2\eta _{02}^{2}\kappa ^{2}\right) e^{\left( \kappa
^{2}\left( 1+\frac{1}{\rho ^{2}}\right) \eta _{02}^{2}\right) }  \nonumber \\
&&-\frac{8\Upsilon \left( Z\right) }{\rho }A_{1}^{2}b^{2}\kappa ^{3}\eta
_{02}^{5}e^{\left( \kappa ^{2}\left( 1+\frac{1}{\rho ^{2}}\right) \eta
_{02}^{2}\right) }  \nonumber \\
&&-\frac{4\Upsilon \left( Z\right) }{\rho }A_{1}^{2}b^{2}\kappa ^{3}\eta
_{02}^{5}\left( 1+2\eta _{02}^{2}\kappa ^{2}\right) \left( 1+\frac{1}{\rho
^{2}}\right) e^{\left( \kappa ^{2}\left( 1+\frac{1}{\rho ^{2}}\right) \eta
_{02}^{2}\right) }.  \TCItag{42d}
\end{eqnarray}
The eigenvalues of this system (eq.(41)) determine the stability near each
of the fixed points and are given by

\begin{eqnarray}
\lambda _{1,2} &=&\pm \sqrt{\left( M_{31}M_{13}\right) },  \TCItag{43a} \\
\lambda _{3,4} &=&\pm \sqrt{\left( M_{42}M_{24}\right) }  \TCItag{43b}
\end{eqnarray}
The eqs.(43) show that the fixed points \textbf{II} and \textbf{III} ($%
\left( \eta _{1},\beta _{1}=0\right) $, $\left( \eta _{2},\beta
_{1}=0\right) $, $\left( \eta _{1},\beta _{2}=0\right) $ and $\left( \eta
_{2},\beta _{2}=0\right) $) are the\textbf{\ }centers if the sign of the
dispersion $\left( \sigma \right) $ is positive. We will then investigate
both the anomalous- and normal-dispersion fibers in what follows.

\textbf{A}. \textbf{Normal Dispersion:} $\left( \sigma _{1}=\sigma
_{2}=\sigma \right) =\frac{D_{-}}{\overline{D}}<0$

\bigskip We start by considering the phases-plane dynamics in the normal
dispersion regime. In this case, we can rewrite $\sigma =-\frac{\left|
D_{-}\right| }{\overline{D}}<0.$ We then consider the location and stability
of the three fixed points of eqs.(23) shown in eqs.(38). Fixed point \textbf{%
I }has already been determined to be generate, and fixed points \textbf{II}
and \textbf{III} lay in the left half-plane since $\eta _{0_{j}}<0.$ Because
we consider only values of $\eta _{j}\geq 0,$ only the fixed points at the
origin are relevant. the shapes of the plane are the same as homoclinic
orbits, which emanate and terminate in the origin.

\textbf{B}. \textbf{Anomalous Dispersion:} $\left( \sigma _{1}=\sigma
_{2}=\sigma \right) =\frac{D_{+}}{\overline{D}}>0$

In the anomalous-dispersion regime, the phases-plane dynamics are
significantly different than that of the normal-dispersion regime since
fixed points \textbf{II} and \textbf{III} , given by eqs.(38b) and (38c) are
located at 
\begin{equation}
\mathbf{II}.\text{ \ \ \ } 
\begin{array}{c}
\beta _{j}=0,\text{ \ and} \\ 
\\ 
\frac{\frac{\Upsilon \left( Z\right) }{\sqrt{2}}A^{2}b^{2}\kappa ^{2}\left(
1+\rho ^{2}\right) }{\frac{2\kappa ^{4}\left( 1+\rho ^{2}\right) D_{+}}{%
\overline{D}}-2\kappa ^{2}\exp \left( \vartheta _{1}\right) \left[ 2\Upsilon
\left( Z\right) \rho A^{2}b^{2}\kappa \right] \vartheta _{1}-\kappa
^{2}\left( 1+\rho ^{2}\right) \exp \left( \vartheta _{1}\right) \left[
2\Upsilon \left( Z\right) \rho A^{2}b^{2}\kappa \right] }{\huge >}0,
\end{array}
\tag{44a}
\end{equation}
and

\begin{equation}
\mathbf{III.}\text{ \ } 
\begin{array}{c}
\beta _{j}=0,\text{ \ and} \\ 
\\ 
\frac{\frac{\Upsilon \left( Z\right) }{\sqrt{2}}A^{2}b^{2}\kappa ^{2}\left(
1+\frac{1}{\rho ^{2}}\right) }{\frac{2\kappa ^{4}\left( 1+\frac{1}{\rho ^{2}}%
\right) D_{+}}{\overline{D}}-2\kappa ^{2}\exp \left( \vartheta _{1}^{\prime
}\right) \left[ 2\Upsilon \left( Z\right) \rho A^{2}b^{2}\kappa \right]
\vartheta _{1}^{\prime }-\kappa ^{2}\left( 1+\frac{1}{\rho ^{2}}\right) \exp
\left( \vartheta _{1}^{\prime }\right) \left[ 2\Upsilon \left( Z\right) \rho
A^{2}b^{2}\kappa \right] }{\Huge >}0,
\end{array}
\text{\ \ }  \tag{44b}
\end{equation}
with the four eigenvalues $\left( \text{for \ \ }\sigma =\frac{D_{+}}{%
\overline{D}}\right) $

\begin{eqnarray}
\lambda _{1,2} &=&\pm \sqrt{\left( M_{31}M_{13}\right) },  \TCItag{45a} \\
\lambda _{3,4} &=&\pm \sqrt{\left( M_{42}M_{24}\right) }  \TCItag{45b}
\end{eqnarray}

In this case, the phases flow for $\eta _{j}\geq 0$ are partially determined
by the three critical points \textbf{I}, \textbf{II} and \textbf{III }. It
can be more obviously understood if we replace $C=-\left( \frac{d\phi
_{1}\left( Z\right) }{dZ}+\frac{d\phi _{2}\left( Z\right) }{dZ}\right) $ in
the eqs.(32) and $r_{a_{j}}=\exp \left( -0.5i\phi _{j}\left( Z\right)
\right) $ in eq.(30e), We then give rise to the separatix

\begin{equation}
\beta _{1}=\pm \eta _{1}\left\{ \gamma _{a}\eta _{1}+\gamma _{c}\eta
_{1}^{2}\exp \left[ \kappa ^{2}\eta _{1}^{2}\left( 1+\rho ^{2}\right) \right]
+\frac{\gamma _{5}}{2}-\left( \gamma _{b}+\frac{1}{2}\right) \eta
_{1}^{2}\right\} ^{\frac{1}{2}},  \tag{46a}
\end{equation}

\begin{eqnarray}
\beta _{1} &=&\pm \eta _{2}\left\{ \frac{\gamma _{a}}{\rho ^{3}}\eta _{2}+%
\frac{\gamma _{c}}{\rho ^{4}}\eta _{2}^{2}\left( e^{\kappa ^{2}\left( 1+%
\frac{1}{\rho ^{2}}\right) \eta _{2}^{2}}\right) +\frac{\gamma _{5}}{2\rho
^{2}}-\left( \frac{1+2\gamma _{b}}{2\rho ^{4}}\right) \eta _{2}^{2}\right\}
^{\frac{1}{2}},  \nonumber \\
&&  \TCItag{46b}
\end{eqnarray}

\begin{eqnarray}
\beta _{2} &=&\pm \eta _{2}\left\{ \frac{\gamma _{a}\eta _{2}}{\rho }+\frac{%
\gamma _{c}\eta _{2}^{2}\left( e^{\kappa ^{2}\left( 1+\frac{1}{\rho ^{2}}%
\right) \eta _{2}^{2}}\right) }{\rho ^{2}}+\frac{\gamma _{5}}{2}-\left( 
\frac{1+2\gamma _{b}}{2\rho ^{2}}\right) \eta _{2}^{2}\right\} ^{\frac{1}{2}%
},  \nonumber \\
&&  \TCItag{46c}
\end{eqnarray}
and

\begin{eqnarray}
\beta _{2} &=&\pm \eta _{1}\left\{ \rho ^{2}\gamma _{a}\eta _{1}+\rho
^{2}\gamma _{c}\eta _{1}^{2}\left( e^{\kappa ^{2}\eta _{1}^{2}\left( 1+\rho
^{2}\right) }\right) +\frac{\rho ^{2}\gamma _{5}}{2}-\left( \rho ^{2}\gamma
_{b}+\frac{\rho ^{2}}{2}\right) \eta _{1}^{2}\right\} ^{\frac{1}{2}}, 
\nonumber \\
&&  \TCItag{46d}
\end{eqnarray}
which have a cusp at the origin $\left( \beta _{j}=\eta _{j}=0\right) $. The
solutions outside these separatix eventually flow into the origin
(homoclinic orbits), while those inside the separatix are periodic. In
eqs.(46), there is an indicator $\gamma _{5}$ represented the contribution
of a perturbation term caused by a noise sourced by amplifiers in optical
fiber and it can influence the amplitudes and chirps of the pulses dynamics.
The geometrical figures and its analysis of this system of eqs.(23) are not
provided in this paper due to a complicated interpretation. We will then
show the simplest explanation of the unperturbed dynamical system in the
following section.

\section{The Hamiltonian Dynamics of The Unperturbed Coupled NLS}

\bigskip

The solution of $\beta _{1}$ in eq.(29) can be reduced to that in \textbf{%
the single} \textbf{NLS} which has no the perturbation terms provided by
Kutz, et al. by setting $\beta _{2}=\eta _{2}=\phi _{2}=0$, $\ \Upsilon
\left( Z\right) =1$ and $\epsilon _{j}R_{j}=0$. On the other hand, we
conclude that the terms of the Lagrangian, the \textbf{ODEs} and the $\beta $
solutions of the unperturbed coupled \textbf{NLS} :

\begin{equation}
\begin{array}{c}
i\frac{\partial U_{1}}{\partial Z}+\frac{\sigma \left( Z\right) }{2}\frac{%
\partial ^{2}U_{1}}{\partial T^{2}}+\left( \left| U_{1}\right| ^{2}+\left|
U_{2}\right| ^{2}\right) U_{1}=0 \\ 
i\frac{\partial U_{2}}{\partial Z}+\frac{\sigma \left( Z\right) }{2}\frac{%
\partial ^{2}U_{2}}{\partial T^{2}}+\left( \left| U_{1}\right| ^{2}+\left|
U_{2}\right| ^{2}\right) U_{2}=0
\end{array}
\tag{47}
\end{equation}
reduced from the results in section 5\textbf{\ }are, respectively, as follows

\begin{eqnarray}
L &=&\sum\limits_{j=1}^{2}\frac{A_{j}^{2}b_{j}^{2}}{\kappa _{j}}\left\{ 
\frac{\sqrt{\frac{\pi }{2}}}{2\eta _{j}^{2}}\frac{\partial \beta _{j}}{%
\partial Z}+\left( \sqrt{\frac{\pi }{2}}\right) \frac{\partial \phi
_{j}\left( Z\right) }{\partial Z}-\frac{\sqrt{\pi }\eta _{j}}{2}\left(
A_{j}^{2}b_{j}^{2}\right) \right\}  \nonumber \\
&&+\sum\limits_{j=1}^{2}\frac{A_{j}^{2}b_{j}^{2}}{\kappa _{j}}\left\{ \sqrt{%
\frac{\pi }{2}}\sigma _{j}\left( Z\right) \kappa _{j}^{2}\left[ \eta
_{j}^{2}+\left( \frac{\beta _{j}^{2}}{\eta _{j}^{2}}\right) \right] \right\}
\nonumber \\
&&-2\sqrt{\frac{\pi }{2}}\eta _{1}\eta
_{2}A_{1}^{2}A_{2}^{2}b_{1}^{2}b_{2}^{2}\left( \exp \left[ \left( \kappa
_{1}^{2}\eta _{1}^{2}+\kappa _{2}^{2}\eta _{2}^{2}\right) \right] \right) , 
\TCItag{48a}
\end{eqnarray}
\begin{eqnarray}
\frac{\partial \beta _{1}}{\partial Z} &=&2\sigma _{1}\left( Z\right) \kappa
_{1}^{2}\left[ \eta _{1}^{4}-\beta _{1}^{2}\right] -\frac{1}{\sqrt{2}}\eta
_{1}^{3}A_{1}^{2}b_{1}^{2}  \nonumber \\
&&-2\eta _{2}\eta _{1}^{3}A_{2}^{2}b_{2}^{2}\kappa _{1}\left( 1+2\eta
_{1}^{2}\kappa _{1}^{2}\right) \exp \left( \kappa _{1}^{2}\eta
_{1}^{2}+\kappa _{2}^{2}\eta _{2}^{2}\right)  \TCItag{48b}
\end{eqnarray}

\begin{eqnarray}
\frac{\partial \beta _{2}}{\partial Z} &=&2\sigma _{2}\left( Z\right) \kappa
_{2}^{2}\left[ \eta _{2}^{4}-\beta _{2}^{2}\right] -\frac{1}{\sqrt{2}}\eta
_{2}^{3}A_{2}^{2}b_{2}^{2}  \nonumber \\
&&-2\eta _{1}\eta _{2}^{3}A_{1}^{2}b_{1}^{2}\kappa _{2}\left( 1+2\eta
_{2}^{2}\kappa _{2}^{2}\right) \exp \left( \kappa _{1}^{2}\eta
_{1}^{2}+\kappa _{2}^{2}\eta _{2}^{2}\right) ,  \TCItag{48c}
\end{eqnarray}

\begin{equation}
\frac{d\eta _{j}}{dZ}=-b_{j}^{2}\sqrt{\frac{\pi }{2}}\sigma _{j}\left(
Z\right) \kappa _{j}\beta _{j}\eta _{j},\text{ \ \ \ \ }j=1,2  \tag{48d}
\end{equation}
and 
\begin{equation}
\beta _{1}=\pm \eta _{1}\left\{ \gamma _{1}\left[ \eta _{1}+\eta _{2}\right]
-\gamma _{2}-\gamma _{3}\left( \eta _{2}^{2}+\frac{\beta _{2}^{2}}{\eta
_{2}^{2}}\right) +\gamma _{4}\eta _{1}\eta _{2}e^{\left[ \kappa ^{2}\left(
\eta _{1}^{2}+\eta _{2}^{2}\right) \right] }-\eta _{1}^{2}\right\} ^{\frac{1%
}{2}}.  \tag{49}
\end{equation}
The simplest stable solution of the equations for the amplitudes $\eta _{j}$
and chirps $\beta _{j}$ derived from eq.(49) will be found by choosing $\eta
_{2}=\rho \eta _{1},$ and $\ \beta _{2}=\rho \beta _{1},$%
\begin{equation}
\beta _{1}=\pm \frac{\eta _{1}}{\sqrt{2}}\left\{ \gamma _{1}\left[ 1+\rho %
\right] \eta _{1}-\gamma _{2}-\rho ^{2}\eta _{1}^{2}+\gamma _{4}\rho \eta
_{1}^{2}e^{\left[ \kappa ^{2}\left( 1+\rho ^{2}\right) \eta _{1}^{2}\right]
}-\eta _{1}^{2}\right\} ^{\frac{1}{2}},  \tag{50a}
\end{equation}
\begin{equation}
\beta _{1}=\pm \frac{\eta _{2}}{\rho \sqrt{2}}\left\{ \gamma _{1}\left[
1+\rho \right] \frac{\eta _{2}}{\rho }-\gamma _{2}-\eta _{2}^{2}+\gamma _{4}%
\frac{\eta _{2}^{2}}{\rho }e^{\left[ \frac{\kappa ^{2}\left( 1+\rho
^{2}\right) }{\rho ^{2}}\eta _{2}^{2}\right] }-\frac{\eta _{2}^{2}}{\rho ^{2}%
}\right\} ^{\frac{1}{2}},  \tag{50b}
\end{equation}
\begin{equation}
\beta _{2}=\pm \left( \frac{\rho }{\sqrt{2}}\right) \eta _{1}\left\{ \gamma
_{1}\left[ 1+\rho \right] \eta _{1}-\gamma _{2}-\rho ^{2}\eta
_{1}^{2}+\gamma _{4}\rho \eta _{1}^{2}e^{\left[ \kappa ^{2}\left( 1+\rho
^{2}\right) \eta _{1}^{2}\right] }-\eta _{1}^{2}\right\} ^{\frac{1}{2}}, 
\tag{50c}
\end{equation}
and 
\begin{equation}
\beta _{2}=\pm \frac{\eta _{2}}{\sqrt{2}}\left\{ \gamma _{1}\left[ 1+\rho %
\right] \frac{\eta _{2}}{\rho }-\gamma _{2}-\eta _{2}^{2}+\gamma _{4}\frac{%
\eta _{2}^{2}}{\rho }e^{\left[ \frac{\kappa ^{2}\left( 1+\rho ^{2}\right) }{%
\rho ^{2}}\eta _{2}^{2}\right] }-\frac{\eta _{2}^{2}}{\rho ^{2}}\right\} ^{%
\frac{1}{2}},  \tag{50d}
\end{equation}
where $\left( \text{for \ }\sigma _{1}\left( Z\right) =\sigma _{2}\left(
Z\right) =\sigma \left( Z\right) \text{ and }b_{1}=b_{2}=b\right) $ 
\begin{eqnarray}
\gamma _{1} &=&\left( \frac{2}{\sigma \left( Z\right) \kappa \sqrt{\pi }}%
\right) ,  \TCItag{51a} \\
\gamma _{2} &=&\left( \frac{C+\frac{d\phi _{1}\left( Z\right) }{dZ}+\frac{%
d\phi _{2}\left( Z\right) }{dZ}}{\sigma \left( Z\right) \kappa ^{2}}\right) ,
\TCItag{51b} \\
\gamma _{4} &=&\frac{4\sqrt{2}}{\sigma \left( Z\right) \sqrt{\pi }}. 
\TCItag{51c}
\end{eqnarray}

The substitution of eqs.(50a) and (50d) into eq.(48d) yields the following
first \textbf{ODEs }in terms of $\eta _{1}$ and $\eta _{2}$ , respectively: 
\begin{equation}
\begin{array}{c}
\frac{d\eta _{1}}{dZ}=\mp \frac{b^{2}\sqrt{\pi }\sigma \left( Z\right)
\kappa \eta _{1}^{2}}{2}\left\{ \gamma _{1}\left[ 1+\rho \right] \eta
_{1}-\gamma _{2}-\rho ^{2}\eta _{1}^{2}+\gamma _{4}\rho \eta _{1}^{2}e^{%
\left[ \kappa ^{2}\left( 1+\rho ^{2}\right) \eta _{1}^{2}\right] }-\eta
_{1}^{2}\right\} ^{\frac{1}{2}},
\end{array}
\tag{52a}
\end{equation}
and 
\begin{equation}
\begin{array}{c}
\frac{d\eta _{2}}{dZ}=\mp \frac{b^{2}\sqrt{\pi }\sigma \left( Z\right)
\kappa \eta _{2}^{2}}{2}\left\{ \gamma _{1}\left[ 1+\rho \right] \frac{\eta
_{2}}{\rho }-\gamma _{2}-\eta _{2}^{2}+\frac{\gamma _{4}\eta _{2}^{2}}{\rho }%
e^{\left[ \frac{\kappa ^{2}\left( 1+\rho ^{2}\right) }{\rho ^{2}}\eta
_{2}^{2}\right] }-\frac{\eta _{2}^{2}}{\rho ^{2}}\right\} ^{\frac{1}{2}}.
\end{array}
\tag{52b}
\end{equation}
Upon integrating these equations with initial conditions $\eta _{j}=\eta
_{0_{j}}$ at $Z=0$, we find

\begin{eqnarray}
Z &=&\mp 2\int\limits_{\eta _{0_{1}}}^{\eta _{1}}\frac{\left\{ \gamma _{1}%
\left[ 1+\rho \right] \eta _{1}-\gamma _{2}-\rho ^{2}\eta _{1}^{2}+\gamma
_{4}\rho \eta _{1}^{2}e^{\left[ \kappa ^{2}\left( 1+\rho ^{2}\right) \eta
_{1}^{2}\right] }-\eta _{1}^{2}\right\} ^{-\frac{1}{2}}d\eta _{1}}{b^{2}%
\sqrt{\pi }\sigma \left( Z\right) \kappa \eta _{1}^{2}},  \nonumber \\
&&  \TCItag{53a}
\end{eqnarray}
or

\begin{eqnarray}
Z &=&\mp 2\int\limits_{\eta _{02}}^{\eta _{2}}\frac{\left\{ \gamma _{1}\left[
1+\frac{1}{\rho }\right] \eta _{2}-\gamma _{2}-\eta _{2}^{2}+\gamma _{4}%
\frac{\eta _{2}^{2}}{\rho }\exp \left[ \frac{\kappa ^{2}\left( 1+\rho
^{2}\right) \eta _{2}^{2}}{\rho ^{2}}\right] -\frac{\eta _{2}^{2}}{\rho ^{2}}%
\right\} ^{-\frac{1}{2}}d\eta _{1}}{b^{2}\sqrt{\pi }\sigma \left( Z\right)
\kappa \eta _{2}^{2}},  \nonumber \\
&&  \TCItag{53b}
\end{eqnarray}
where $\sigma \left( Z\right) $ is a fixed value, i.e., $\sigma _{1}=\sigma
_{2}=\frac{D_{-}}{\overline{D}}$ or $\frac{D_{+}}{\overline{D}}$, depending
on whether the coupled pulses are propagating in the normal or anomalous
regime, respectively. The integrations in eq.(53) are very complicated
expressions, which must be transformed to give $\eta _{j}=\eta _{j}\left(
Z\right) $. However, we will explain it elsewhere. The only simplest manner
in describing the coupled breathers dynamics is by using a geometrical
interpretation of eqs.( 48b), (48c) and (48d).

The fixed points derived from a perturbative coupled \textbf{NLS }are
usually very complicated in terms of the roots. However, in order to explain
the phases-plane dynamics of the coupled breathers, we come to obtain the
unperturbed fixed points of eqs.(48) by setting the right-hand sides equal
to zero. Here, we find three fixed points reduced from that in eqs.(44):

\begin{equation}
\mathbf{I.}\text{ \ \ \ }\beta _{j}=0,\text{ \ \ \ \ \ }\eta _{j}=0, 
\tag{54a}
\end{equation}

\begin{eqnarray}
\mathbf{II}.\text{ \ \ \ }\beta _{j} &=&0,\text{ }  \nonumber \\
\text{\ \ }\eta _{1} &=&\frac{a_{21}a_{71}}{a_{71}a_{11}-a_{41}\exp \left(
\vartheta _{1}\right) a_{31}\vartheta _{1}-a_{71}\exp \left( \vartheta
_{1}\right) a_{31}}=\eta _{10},  \TCItag{54b}
\end{eqnarray}

\begin{eqnarray}
\mathbf{III.}\text{ \ \ \ }\beta _{j} &=&0,\text{\ \ \ }  \nonumber \\
\text{\ \ }\eta _{2} &=&\frac{c_{21}c_{71}}{c_{71}c_{11}-c_{41}\exp \left(
\vartheta _{1}^{\prime }\right) c_{31}\vartheta _{1}^{\prime }-c_{71}\exp
\left( \vartheta _{1}^{\prime }\right) c_{31}}=\eta _{20},  \TCItag{54c}
\end{eqnarray}
\bigskip where $\vartheta _{1}$ is a root of

\begin{eqnarray}
&&{\Huge \{}e^{2\Theta }\Theta a_{71}^{2}a_{31}^{2}+2e^{2\Theta
}a_{71}a_{31}^{2}a_{41}\Theta ^{2}+e^{2\Theta }a_{41}^{2}a_{31}^{2}\Theta
^{3}-2e^{\Theta }\Theta a_{71}^{2}a_{11}a_{31}  \nonumber \\
&&-2e^{\Theta }a_{71}a_{11}a_{41}a_{31}\Theta
^{2}-a_{21}^{2}a_{71}^{3}+\Theta a_{71}^{2}a_{11}^{2}{\Huge \}}, 
\TCItag{55a}
\end{eqnarray}
and $\vartheta _{1}^{\prime }$ is a root of 
\begin{eqnarray}
&&{\Huge \{}e^{2\Theta }\Theta c_{71}^{2}c_{31}^{2}+2e^{2\Theta
}c_{71}c_{31}^{2}c_{41}\Theta ^{2}+e^{2\Theta }c_{41}^{2}c_{31}^{2}\Theta
^{3}-2e^{\Theta }\Theta c_{71}^{2}c_{11}c_{31}  \nonumber \\
&&-2e^{\Theta }c_{71}c_{11}c_{41}c_{31}\Theta
^{2}-c_{21}^{2}c_{71}^{3}+\Theta c_{71}^{2}c_{11}^{2}{\Huge \}}, 
\TCItag{55b}
\end{eqnarray}
in which the roots $\vartheta _{1}$and $\vartheta _{1}^{\prime }$ are,
respectively, $\vartheta _{1}=1.\,\allowbreak 316\,5$, $\ $and $\ \vartheta
_{1}^{\prime }=6.\,\allowbreak 189+3.\,\allowbreak 136i$ \ if $A=1$, $\sigma
=1$, and $\rho =0.03$, and the parameters $\eta _{01},\eta
_{02},a_{11},a_{21}$,$a_{31},a_{41}$, $a_{71}$, $c_{11},c_{21}$,$%
c_{31},c_{41},$and $c_{71}$ ($A_{1}=A_{2}=A$) are

\begin{equation}
\begin{array}{c}
\eta _{01}=\frac{\frac{1}{\sqrt{2}}A^{2}b^{2}\kappa ^{2}\left( 1+\rho
^{2}\right) }{2\kappa ^{4}\left( 1+\rho ^{2}\right) \sigma -2\kappa ^{2}\exp
\left( \vartheta _{1}\right) \left[ 2\rho A^{2}b^{2}\kappa \right] \vartheta
_{1}-\kappa ^{2}\left( 1+\rho ^{2}\right) \exp \left( \vartheta _{1}\right) 
\left[ 2\rho A^{2}b^{2}\kappa \right] },
\end{array}
\tag{56a}
\end{equation}

\begin{equation}
\begin{array}{c}
\eta _{02}=\frac{\frac{1}{\sqrt{2}}A^{2}b^{2}\kappa ^{2}\left( 1+\frac{1}{%
\rho ^{2}}\right) }{2\kappa ^{4}\left( 1+\frac{1}{\rho ^{2}}\right) \sigma
-2\kappa ^{2}\exp \left( \vartheta _{1}^{\prime }\right) \left[ 2\rho
A^{2}b^{2}\kappa \right] \vartheta _{1}^{\prime }-\kappa ^{2}\left( 1+\frac{1%
}{\rho ^{2}}\right) \exp \left( \vartheta _{1}^{\prime }\right) \left[ 2\rho
A^{2}b^{2}\kappa \right] },
\end{array}
\tag{56b}
\end{equation}
\begin{eqnarray*}
a_{11} &=&c_{11}=2\sigma \left( Z\right) \kappa ^{2}, \\
a_{21} &=&c_{21}=\frac{A^{2}b^{2}}{\sqrt{2}}, \\
a_{31} &=&c_{31}=2\rho A^{2}b^{2}\kappa , \\
a_{41} &=&c_{41}=2\kappa ^{2}, \\
a_{71} &=&\kappa ^{2}\left( 1+\rho ^{2}\right) , \\
c_{71} &=&\kappa ^{2}\left( 1+\frac{1}{\rho ^{2}}\right) .
\end{eqnarray*}

Linearizing of eqs.(48b), (48c) and (48d) by setting $\eta _{j}=\eta _{0j}+%
\widetilde{\eta }_{j}$ and $\beta _{j}=0+\widetilde{\beta _{j}}$, we find
the linear system by replacing $\epsilon _{j}R_{j}=0$ or $r_{a_{1}}=0,$ and $%
\Upsilon \left( Z\right) =1$ in eqs.(42) and it can be rewritten in a matrix
form as follows

\begin{equation}
\left( 
\begin{array}{c}
\frac{d\widetilde{\eta _{1}}}{dZ} \\ 
\frac{d\widetilde{\eta _{2}}}{dZ} \\ 
\frac{\partial \widetilde{\beta _{1}}}{\partial Z} \\ 
\frac{\partial \widetilde{\beta _{2}}}{\partial Z}
\end{array}
\right) =\left( 
\begin{array}{cccc}
0 & 0 & M_{13}^{\prime } & 0 \\ 
0 & 0 & 0 & M_{24}^{\prime } \\ 
M_{31}^{\prime } & 0 & 0 & 0 \\ 
0 & M_{42}^{\prime } & 0 & 0
\end{array}
\right) \left( 
\begin{array}{c}
\widetilde{\eta _{1}} \\ 
\widetilde{\eta _{2}} \\ 
\widetilde{\beta _{1}} \\ 
\widetilde{\beta _{2}}
\end{array}
\right) ,  \tag{57}
\end{equation}
where 
\begin{equation}
M_{13}^{\prime }=-b^{2}\sqrt{\frac{\pi }{2}}\sigma \left( Z\right) \kappa
\eta _{01}  \tag{58a}
\end{equation}

\begin{equation}
M_{24}^{\prime }=-b^{2}\sqrt{\frac{\pi }{2}}\sigma \left( Z\right) \kappa
\eta _{02}  \tag{58b}
\end{equation}

\begin{eqnarray}
M_{31}^{\prime } &=&8\sigma \left( Z\right) \kappa ^{2}\eta _{01}^{3}-\frac{3%
}{\sqrt{2}}A^{2}b^{2}\eta _{01}^{2}  \nonumber \\
&&-8\rho A^{2}b^{2}\kappa \eta _{01}^{3}\left( 1+2\eta _{01}^{2}\kappa
^{2}\right) e^{\left( \kappa ^{2}\left( 1+\rho ^{2}\right) \eta
_{01}^{2}\right) }  \nonumber \\
&&-8\rho A^{2}b^{2}\kappa ^{3}\eta _{01}^{5}e^{\left( \kappa ^{2}\left(
1+\rho ^{2}\right) \eta _{01}^{2}\right) }  \nonumber \\
&&-4\rho A^{2}b^{2}\kappa ^{3}\eta _{01}^{5}\left( 1+2\eta _{01}^{2}\kappa
^{2}\right) \left( 1+\rho ^{2}\right) e^{\left( \kappa ^{2}\left( 1+\rho
^{2}\right) \eta _{01}^{2}\right) }  \TCItag{58c}
\end{eqnarray}

\begin{eqnarray}
M_{42}^{\prime } &=&8\sigma \left( Z\right) \kappa ^{2}\eta _{02}^{3}-\frac{3%
}{\sqrt{2}}A^{2}b^{2}\eta _{02}^{2}  \nonumber \\
&&-\frac{8}{\rho }A^{2}b^{2}\kappa \eta _{02}^{3}\left( 1+2\eta
_{02}^{2}\kappa ^{2}\right) e^{\left( \kappa ^{2}\left( 1+\frac{1}{\rho ^{2}}%
\right) \eta _{02}^{2}\right) }  \nonumber \\
&&-\frac{8}{\rho }A^{2}b^{2}\kappa ^{3}\eta _{02}^{5}e^{\left( \kappa
^{2}\left( 1+\frac{1}{\rho ^{2}}\right) \eta _{02}^{2}\right) }  \nonumber \\
&&-\frac{4}{\rho }A^{2}b^{2}\kappa ^{3}\eta _{02}^{5}\left( 1+2\eta
_{02}^{2}\kappa ^{2}\right) \left( 1+\frac{1}{\rho ^{2}}\right) e^{\left(
\kappa ^{2}\left( 1+\frac{1}{\rho ^{2}}\right) \eta _{02}^{2}\right) }. 
\TCItag{58d}
\end{eqnarray}
The eigenvalues of the system (eq.(57)) determine the stability near each of
the fixed points and are given by 
\begin{equation}
\begin{array}{cc}
\lambda _{1,2}=\pm {\Huge \{} & -8\sigma ^{2}b^{2}\sqrt{\frac{\pi }{2}}%
\kappa ^{3}\eta _{01}^{4}+\frac{\left[ \frac{12}{\sqrt{2}}\sigma ^{2}b^{2}%
\sqrt{\frac{\pi }{2}}\kappa ^{3}\eta _{01}^{4}\right] \left( 1+\rho
^{2}\right) }{8\eta _{01}\kappa e^{\vartheta _{1}}\rho \vartheta _{1}+4\eta
_{01}\kappa e^{\vartheta _{1}}\rho +4\eta _{01}\kappa e^{\vartheta _{1}}\rho
^{3}+\sqrt{2}+\sqrt{2}\rho ^{2}} \\ 
& +\frac{\left[ 32\sqrt{\frac{\pi }{2}}\rho \sigma ^{2}b^{2}\kappa ^{4}\eta
_{01}^{5}\left( 1+2\eta _{01}^{2}\kappa ^{2}\right) e^{\left( \kappa
^{2}\left( 1+\rho ^{2}\right) \eta _{01}^{2}\right) }\right] \left( 1+\rho
^{2}\right) }{\left( 8\eta _{01}\kappa e^{\vartheta _{1}}\rho \vartheta
_{1}+4\eta _{01}\kappa e^{\vartheta _{1}}\rho +4\eta _{01}\kappa
e^{\vartheta _{1}}\rho ^{3}+\sqrt{2}+\sqrt{2}\rho ^{2}\right) } \\ 
& +\frac{\left[ 32\sqrt{\frac{\pi }{2}}\rho \sigma ^{2}b^{2}\kappa ^{6}\eta
_{01}^{7}e^{\left( \kappa ^{2}\left( 1+\rho ^{2}\right) \eta
_{01}^{2}\right) }\right] \left( 1+\rho ^{2}\right) }{\left( 8\eta
_{01}\kappa e^{\vartheta _{1}}\rho \vartheta _{1}+4\eta _{01}\kappa
e^{\vartheta _{1}}\rho +4\eta _{01}\kappa e^{\vartheta _{1}}\rho ^{3}+\sqrt{2%
}+\sqrt{2}\rho ^{2}\right) } \\ 
& +\frac{\left[ 32\sqrt{\frac{\pi }{2}}\rho \sigma ^{2}b^{2}\kappa ^{6}\eta
_{01}^{7}e^{\left( \kappa ^{2}\left( 1+\rho ^{2}\right) \eta
_{01}^{2}\right) }\right] \left( 1+\rho ^{2}\right) }{\left( 8\eta
_{01}\kappa e^{\vartheta _{1}}\rho \vartheta _{1}+4\eta _{01}\kappa
e^{\vartheta _{1}}\rho +4\eta _{01}\kappa e^{\vartheta _{1}}\rho ^{3}+\sqrt{2%
}+\sqrt{2}\rho ^{2}\right) } \\ 
& +\frac{\left[ 16\sqrt{\frac{\pi }{2}}\rho \sigma ^{2}b^{2}\kappa ^{6}\eta
_{01}^{7}\left( 1+2\eta _{01}^{2}\kappa ^{2}\right) \left( 1+\rho
^{2}\right) e^{\left( \kappa ^{2}\left( 1+\rho ^{2}\right) \eta
_{01}^{2}\right) }\right] \left( 1+\rho ^{2}\right) }{\left( 8\eta
_{01}\kappa e^{\vartheta _{1}}\rho \vartheta _{1}+4\eta _{01}\kappa
e^{\vartheta _{1}}\rho +4\eta _{01}\kappa e^{\vartheta _{1}}\rho ^{3}+\sqrt{2%
}+\sqrt{2}\rho ^{2}\right) }{\Huge \}}^{\frac{1}{2}}
\end{array}
\tag{59a}
\end{equation}

and\bigskip 
\begin{equation}
\begin{array}{cc}
\lambda _{3,4}=\pm {\Huge \{} & -8\sigma ^{2}b^{2}\sqrt{\frac{\pi }{2}}%
\kappa ^{3}\eta _{02}^{4}+\frac{\left[ \frac{12}{\sqrt{2}}\sigma ^{2}b^{2}%
\sqrt{\frac{\pi }{2}}\kappa ^{3}\eta _{02}^{4}\right] \left( 1+\rho
^{2}\right) }{8\eta _{02}\kappa e^{\vartheta _{1}}\rho \vartheta _{1}+4\eta
_{02}\kappa e^{\vartheta _{1}}\rho +4\eta _{02}\kappa e^{\vartheta _{1}}\rho
^{3}+\sqrt{2}+\sqrt{2}\rho ^{2}} \\ 
& +\frac{\left[ 32\sqrt{\frac{\pi }{2}}\sigma ^{2}b^{2}\kappa ^{4}\eta
_{02}^{5}\left( 1+2\eta _{02}^{2}\kappa ^{2}\right) e^{\left( \kappa
^{2}\left( 1+\frac{1}{\rho ^{2}}\right) \eta _{02}^{2}\right) }\right]
\left( 1+\rho ^{2}\right) }{\rho \left( 8\eta _{02}\kappa e^{\vartheta
_{1}}\rho \vartheta _{1}+4\eta _{02}\kappa e^{\vartheta _{1}}\rho +4\eta
_{02}\kappa e^{\vartheta _{1}}\rho ^{3}+\sqrt{2}+\sqrt{2}\rho ^{2}\right) }
\\ 
& +\frac{\left[ 32\sqrt{\frac{\pi }{2}}\sigma ^{2}b^{2}\kappa ^{4}\eta
_{02}^{5}\left( 1+2\eta _{02}^{2}\kappa ^{2}\right) e^{\left( \kappa
^{2}\left( 1+\frac{1}{\rho ^{2}}\right) \eta _{02}^{2}\right) }\right]
\left( 1+\rho ^{2}\right) }{\rho \left( 8\eta _{02}\kappa e^{\vartheta
_{1}}\rho \vartheta _{1}+4\eta _{02}\kappa e^{\vartheta _{1}}\rho +4\eta
_{02}\kappa e^{\vartheta _{1}}\rho ^{3}+\sqrt{2}+\sqrt{2}\rho ^{2}\right) }
\\ 
& +\frac{\left[ 32\sqrt{\frac{\pi }{2}}\sigma ^{2}b^{2}\kappa ^{6}\eta
_{02}^{7}e^{\left( \kappa ^{2}\left( 1+\frac{1}{\rho ^{2}}\right) \eta
_{02}^{2}\right) }\right] \left( 1+\rho ^{2}\right) }{\rho \left( 8\eta
_{02}\kappa e^{\vartheta _{1}}\rho \vartheta _{1}+4\eta _{02}\kappa
e^{\vartheta _{1}}\rho +4\eta _{02}\kappa e^{\vartheta _{1}}\rho ^{3}+\sqrt{2%
}+\sqrt{2}\rho ^{2}\right) } \\ 
& +\frac{\left[ 16\sqrt{\frac{\pi }{2}}\sigma ^{2}b^{2}\kappa ^{6}\eta
_{02}^{7}\left( 1+2\eta _{02}^{2}\kappa ^{2}\right) \left( 1+\frac{1}{\rho
^{2}}\right) e^{\left( \kappa ^{2}\left( 1+\frac{1}{\rho ^{2}}\right) \eta
_{02}^{2}\right) }\right] \left( 1+\rho ^{2}\right) }{\rho \left( 8\eta
_{02}\kappa e^{\vartheta _{1}}\rho \vartheta _{1}+4\eta _{02}\kappa
e^{\vartheta _{1}}\rho +4\eta _{02}\kappa e^{\vartheta _{1}}\rho ^{3}+\sqrt{2%
}+\sqrt{2}\rho ^{2}\right) }{\Huge \}}^{\frac{1}{2}}
\end{array}
\tag{59b}
\end{equation}

The eqs.(59) show that the fixed points \textbf{II} and \textbf{III} are 
\textbf{the centers}, regardless of the sign of the dispersion. We will then
investigate both the anomalous- and normal-dispersion fibers in what follows.

\textbf{A}. \textbf{Normal Dispersion:} $\sigma =\frac{D_{-}}{\overline{D}}%
<0 $

\bigskip\ In this case, we can rewrite $\sigma =-\frac{\left| D_{-}\right| }{%
\overline{D}}<0.$ We then consider the location and stability of the three
fixed points of eqs.(48b), (48c), and (48d) shown in eqs.(54). Fixed point 
\textbf{I }has already been determined to be degenerate, and fixed points 
\textbf{II} and \textbf{III} lay in the left half-plane since $\eta
_{0_{j}}<0.$ Because we consider only values of $\eta _{j}\geq 0,$ only the
fixed points at the origin are relevant. The phases-plane dynamics are shown
in figures 1, 3, 5, \ and 7 along with the arrows in phases superposition
space. We find that the shapes of the plane are the same as homoclinic
orbits, which emanate and terminate in the origin.

\textbf{B}. \textbf{Anomalous Dispersion:} $\sigma =\frac{D_{+}}{\overline{D}%
}>0$

\bigskip In the anomalous-dispersion regime, the phases-plane dynamics is
significantly different than that of the normal-dispersion regime since
fixed points \textbf{II} and \textbf{III} , given by eqs.(54b) and (54c) are
located at

\begin{equation}
\mathbf{II}.\text{ \ } 
\begin{array}{c}
\beta _{j}=0, \\ 
\\ 
\frac{\frac{1}{\sqrt{2}}A^{2}b^{2}\kappa ^{2}\left( 1+\rho ^{2}\right) }{%
2\kappa ^{4}\left( 1+\rho ^{2}\right) \frac{D_{+}}{\overline{D}}-2\kappa
^{2}\exp \left( \vartheta _{1}\right) \left[ 2\rho A^{2}b^{2}\kappa \right]
\vartheta _{1}-\kappa ^{2}\left( 1+\rho ^{2}\right) \exp \left( \vartheta
_{1}\right) \left[ 2\rho A^{2}b^{2}\kappa \right] }{\Huge >}0,
\end{array}
\text{\ \ }  \tag{60a}
\end{equation}
and

\begin{equation}
\mathbf{III.}\text{ \ } 
\begin{array}{c}
\beta _{j}=0,\text{ } \\ 
\\ 
\frac{\frac{1}{\sqrt{2}}A^{2}b^{2}\kappa ^{2}\left( 1+\frac{1}{\rho ^{2}}%
\right) }{2\kappa ^{4}\left( 1+\frac{1}{\rho ^{2}}\right) \frac{D_{+}}{%
\overline{D}}-2\kappa ^{2}\exp \left( \vartheta _{1}\right) \left[ 2\rho
A^{2}b^{2}\kappa \right] \vartheta _{1}-\kappa ^{2}\left( 1+\frac{1}{\rho
^{2}}\right) \exp \left( \vartheta _{1}\right) \left[ 2\rho A^{2}b^{2}\kappa 
\right] }{\Huge >}0,
\end{array}
\text{\ \ }  \tag{60b}
\end{equation}
with the four of eigenvalues 
\begin{equation}
\begin{array}{cc}
\lambda _{1,2}=\pm {\Huge \{} & -8\left( \frac{D_{+}}{\overline{D}}\right)
^{2}b^{2}\sqrt{\frac{\pi }{2}}\kappa ^{3}\eta _{01}^{4}+\frac{\left[ \frac{12%
}{\sqrt{2}}\left( \frac{D_{+}}{\overline{D}}\right) ^{2}b^{2}\sqrt{\frac{\pi 
}{2}}\kappa ^{3}\eta _{01}^{4}\right] \left( 1+\rho ^{2}\right) }{8\eta
_{01}\kappa e^{\vartheta _{1}}\rho \vartheta _{1}+4\eta _{01}\kappa
e^{\vartheta _{1}}\rho +4\eta _{01}\kappa e^{\vartheta _{1}}\rho ^{3}+\sqrt{2%
}+\sqrt{2}\rho ^{2}} \\ 
& +\frac{\left[ 32\sqrt{\frac{\pi }{2}}\rho \left( \frac{D_{+}}{\overline{D}}%
\right) ^{2}b^{2}\kappa ^{4}\eta _{01}^{5}\left( 1+2\eta _{01}^{2}\kappa
^{2}\right) e^{\left( \kappa ^{2}\left( 1+\rho ^{2}\right) \eta
_{01}^{2}\right) }\right] \left( 1+\rho ^{2}\right) }{\left( 8\eta
_{01}\kappa e^{\vartheta _{1}}\rho \vartheta _{1}+4\eta _{01}\kappa
e^{\vartheta _{1}}\rho +4\eta _{01}\kappa e^{\vartheta _{1}}\rho ^{3}+\sqrt{2%
}+\sqrt{2}\rho ^{2}\right) } \\ 
& +\frac{\left[ 32\sqrt{\frac{\pi }{2}}\rho \left( \frac{D_{+}}{\overline{D}}%
\right) ^{2}b^{2}\kappa ^{6}\eta _{01}^{7}e^{\left( \kappa ^{2}\left( 1+\rho
^{2}\right) \eta _{01}^{2}\right) }\right] \left( 1+\rho ^{2}\right) }{%
\left( 8\eta _{01}\kappa e^{\vartheta _{1}}\rho \vartheta _{1}+4\eta
_{01}\kappa e^{\vartheta _{1}}\rho +4\eta _{01}\kappa e^{\vartheta _{1}}\rho
^{3}+\sqrt{2}+\sqrt{2}\rho ^{2}\right) } \\ 
& +\frac{\left[ 32\sqrt{\frac{\pi }{2}}\rho \left( \frac{D_{+}}{\overline{D}}%
\right) ^{2}b^{2}\kappa ^{6}\eta _{01}^{7}e^{\left( \kappa ^{2}\left( 1+\rho
^{2}\right) \eta _{01}^{2}\right) }\right] \left( 1+\rho ^{2}\right) }{%
\left( 8\eta _{01}\kappa e^{\vartheta _{1}}\rho \vartheta _{1}+4\eta
_{01}\kappa e^{\vartheta _{1}}\rho +4\eta _{01}\kappa e^{\vartheta _{1}}\rho
^{3}+\sqrt{2}+\sqrt{2}\rho ^{2}\right) } \\ 
& +\frac{\left[ 16\sqrt{\frac{\pi }{2}}\rho \left( \frac{D_{+}}{\overline{D}}%
\right) ^{2}b^{2}\kappa ^{6}\eta _{01}^{7}\left( 1+2\eta _{01}^{2}\kappa
^{2}\right) \left( 1+\rho ^{2}\right) e^{\left( \kappa ^{2}\left( 1+\rho
^{2}\right) \eta _{01}^{2}\right) }\right] \left( 1+\rho ^{2}\right) }{%
\left( 8\eta _{01}\kappa e^{\vartheta _{1}}\rho \vartheta _{1}+4\eta
_{01}\kappa e^{\vartheta _{1}}\rho +4\eta _{01}\kappa e^{\vartheta _{1}}\rho
^{3}+\sqrt{2}+\sqrt{2}\rho ^{2}\right) }{\Huge \}}^{\frac{1}{2}},
\end{array}
\tag{61a}
\end{equation}
and 
\begin{equation}
\begin{array}{cc}
\lambda _{3,4}=\pm {\Huge \{} & -8\left( \frac{D_{+}}{\overline{D}}\right)
^{2}b^{2}\sqrt{\frac{\pi }{2}}\kappa ^{3}\eta _{02}^{4}+\frac{\left[ \frac{12%
}{\sqrt{2}}\left( \frac{D_{+}}{\overline{D}}\right) ^{2}b^{2}\sqrt{\frac{\pi 
}{2}}\kappa ^{3}\eta _{02}^{4}\right] \left( 1+\rho ^{2}\right) }{8\eta
_{02}\kappa e^{\vartheta _{1}}\rho \vartheta _{1}+4\eta _{02}\kappa
e^{\vartheta _{1}}\rho +4\eta _{02}\kappa e^{\vartheta _{1}}\rho ^{3}+\sqrt{2%
}+\sqrt{2}\rho ^{2}} \\ 
& +\frac{\left[ 32\sqrt{\frac{\pi }{2}}\left( \frac{D_{+}}{\overline{D}}%
\right) ^{2}b^{2}\kappa ^{4}\eta _{02}^{5}\left( 1+2\eta _{02}^{2}\kappa
^{2}\right) e^{\left( \kappa ^{2}\left( 1+\frac{1}{\rho ^{2}}\right) \eta
_{02}^{2}\right) }\right] \left( 1+\rho ^{2}\right) }{\rho \left( 8\eta
_{02}\kappa e^{\vartheta _{1}}\rho \vartheta _{1}+4\eta _{02}\kappa
e^{\vartheta _{1}}\rho +4\eta _{02}\kappa e^{\vartheta _{1}}\rho ^{3}+\sqrt{2%
}+\sqrt{2}\rho ^{2}\right) } \\ 
& +\frac{\left[ 32\sqrt{\frac{\pi }{2}}\left( \frac{D_{+}}{\overline{D}}%
\right) ^{2}b^{2}\kappa ^{6}\eta _{02}^{7}e^{\left( \kappa ^{2}\left( 1+%
\frac{1}{\rho ^{2}}\right) \eta _{02}^{2}\right) }\right] \left( 1+\rho
^{2}\right) }{\rho \left( 8\eta _{02}\kappa e^{\vartheta _{1}}\rho \vartheta
_{1}+4\eta _{02}\kappa e^{\vartheta _{1}}\rho +4\eta _{02}\kappa
e^{\vartheta _{1}}\rho ^{3}+\sqrt{2}+\sqrt{2}\rho ^{2}\right) } \\ 
& +\frac{\left[ 32\sqrt{\frac{\pi }{2}}\left( \frac{D_{+}}{\overline{D}}%
\right) ^{2}b^{2}\kappa ^{6}\eta _{02}^{7}e^{\left( \kappa ^{2}\left( 1+%
\frac{1}{\rho ^{2}}\right) \eta _{02}^{2}\right) }\right] \left( 1+\rho
^{2}\right) }{\rho \left( 8\eta _{02}\kappa e^{\vartheta _{1}}\rho \vartheta
_{1}+4\eta _{02}\kappa e^{\vartheta _{1}}\rho +4\eta _{02}\kappa
e^{\vartheta _{1}}\rho ^{3}+\sqrt{2}+\sqrt{2}\rho ^{2}\right) } \\ 
& +\frac{\left[ 16\sqrt{\frac{\pi }{2}}\left( \frac{D_{+}}{\overline{D}}%
\right) ^{2}b^{2}\kappa ^{6}\eta _{02}^{7}\left( 1+2\eta _{02}^{2}\kappa
^{2}\right) \left( 1+\frac{1}{\rho ^{2}}\right) e^{\left( \kappa ^{2}\left(
1+\frac{1}{\rho ^{2}}\right) \eta _{02}^{2}\right) }\right] \left( 1+\rho
^{2}\right) }{\rho \left( 8\eta _{02}\kappa e^{\vartheta _{1}}\rho \vartheta
_{1}+4\eta _{02}\kappa e^{\vartheta _{1}}\rho +4\eta _{02}\kappa
e^{\vartheta _{1}}\rho ^{3}+\sqrt{2}+\sqrt{2}\rho ^{2}\right) }{\Huge \}}^{%
\frac{1}{2}}
\end{array}
\tag{61b}
\end{equation}
The phases-plane dymamics are shown in figures 2, 4, 6, and 8. In this case,
the phases flow for $\eta _{j}\geq 0$ are partially determined by the three
critical points \textbf{I}, \textbf{II} and \textbf{III }. It can be more
obviously understood if we replace 
\[
C=-\left( \frac{d\phi _{1}\left( Z\right) }{dZ}+\frac{d\phi _{2}\left(
Z\right) }{dZ}\right) , 
\]
in the eqs.(50), We then give rise to the separatix 
\begin{equation}
\begin{array}{c}
\beta _{1}=\pm \frac{\eta _{1}}{\sqrt{2}}\left\{ \frac{\left[ 2+2\rho \right]
\overline{D}\eta _{1}}{D_{+}\kappa \sqrt{\pi }}+\frac{4\sqrt{2}\rho 
\overline{D}\eta _{1}^{2}\exp \left[ \kappa ^{2}\left( 1+\rho ^{2}\right)
\eta _{1}^{2}\right] }{D_{+}\sqrt{\pi }}-\left( \rho ^{2}+1\right) \eta
_{1}^{2}\right\} ^{\frac{1}{2}},
\end{array}
\tag{62a}
\end{equation}
\begin{equation}
\begin{array}{c}
\beta _{1}=\pm \frac{\eta _{2}}{\rho \sqrt{2}}\left\{ \frac{\left[ \frac{2}{%
\rho }+2\right] \overline{D}\eta _{2}}{D_{+}\kappa \sqrt{\pi }}+\frac{\frac{%
\overline{D}4\sqrt{2}\eta _{2}^{2}}{\rho }\exp \left[ \kappa ^{2}\left( 
\frac{1}{\rho ^{2}}+1\right) \eta _{2}^{2}\right] }{D_{+}\sqrt{\pi }}-\left(
1+\frac{1}{\rho ^{2}}\right) \eta _{2}^{2}\right\} ^{\frac{1}{2}},
\end{array}
\tag{62b}
\end{equation}
\begin{equation}
\begin{array}{c}
\beta _{2}=\pm \left( \frac{\rho }{\sqrt{2}}\right) \eta _{1}\left\{ \frac{%
\left[ 2+2\rho \right] \overline{D}\eta _{1}}{D_{+}\kappa \sqrt{\pi }}+\frac{%
4\sqrt{2}\rho \overline{D}\eta _{1}^{2}\exp \left[ \kappa ^{2}\left( 1+\rho
^{2}\right) \eta _{1}^{2}\right] }{D_{+}\sqrt{\pi }}-\left( \rho
^{2}+1\right) \eta _{1}^{2}\right\} ^{\frac{1}{2}},
\end{array}
\tag{62c}
\end{equation}
and 
\begin{equation}
\begin{array}{c}
\beta _{2}=\pm \frac{\eta _{2}}{\sqrt{2}}\left\{ \frac{\left[ \frac{2}{\rho }%
+2\right] \overline{D}\eta _{2}}{D_{+}\kappa \sqrt{\pi }}+\frac{\frac{%
\overline{D}4\sqrt{2}\eta _{2}^{2}}{\rho }\exp \left[ \kappa ^{2}\left( 
\frac{1}{\rho ^{2}}+1\right) \eta _{2}^{2}\right] }{D_{+}\sqrt{\pi }}-\left(
1+\frac{1}{\rho ^{2}}\right) \eta _{2}^{2}\right\} ^{\frac{1}{2}},
\end{array}
\tag{62d}
\end{equation}
which have a cusp at the origin $\left( \beta _{j}=\eta _{j}=0\right) $. The
solutions outside these separatix eventually flow into the origin
(homoclinic orbits), while those inside the separatix are periodic.\bigskip\ 

From figures 2, 4, 6, and 8 in which the parameter $\rho $ has been set to
be, for instance, $\rho =0.03$, we then find exactly that the fixed points
of $\left( \eta _{1},\beta _{1}=0\right) $, $\left( \eta _{2},\beta
_{1}=0\right) $, $\left( \eta _{1},\beta _{2}=0\right) $ and $\left( \eta
_{2},\beta _{2}=0\right) $ are at the center of their graphics as follows

\begin{equation}
\left( \text{ }\eta _{1},\left( \beta _{1}=0\right) \right) =\left( \text{ \ 
}\eta _{1},\left( \beta _{2}=0\right) \right) =\left( \text{ \ \ }0.75\text{
\ },\text{ \ }0\text{ \ }\right) ,  \tag{63a}
\end{equation}
and

\begin{equation}
\left( \text{ \ }\eta _{2},\left( \beta _{1}=0\right) \right) =\left( \text{
\ }\eta _{2},\left( \beta _{2}=0\right) \right) =\left( \text{\ }0.0225\text{
}\ ,\text{ }0\text{ }\right) .  \tag{63b}
\end{equation}
The points and the phases-plane dynamics of the \textbf{ODEs} eqs.(48b),
(48c) and (48d) improve that our suggestion in eq.(31c), $\rho \left(
Z\right) =\frac{\beta _{2}}{\beta _{1}}=\frac{\eta _{2}}{\eta _{1}}=const.$,
is actually proved. This means the coupled breathers dynamics depent on the
certain relationship of the chirps and amplitudes. The choice of an
arbitrary $\rho \left( Z\right) $ makes the certain fixed points of $\left(
\eta _{j},\beta _{j}\right) $ obeyed by eqs.(54). These fixed points are
found in $\left( C+\frac{d\phi _{1}\left( Z\right) }{dZ}+\frac{d\phi
_{2}\left( Z\right) }{dZ}\right) \approx 0.605$. In the physical meaning,
the interactions between both chirps depend on a linear constant $\rho
\left( Z\right) $. However, the magnitude of the Gaussian pulses amplitudes
is not only directly influenced by the chirp itself but also by the
interaction of both chrips.

On the other hand, we, for instance, can set the values of $\rho \left(
Z\right) $ and then find the the amplitudes by using a numerical method from
their phases-plane dynamics graphics.

A geometrical representation of the solution and its associated dynamics
behavior of eqs.(48b), (48c) and (48d) appeared in figures 1-8 show the
superposition of the flow of phases $\left( \phi _{1}\text{ and }\phi
_{2}\right) $. According to the figures, the construction of a periodic
dispersion-managed coupled breathers solution from the dispersion map
depends on the critical values of $A_{j}$. In this periodic solutions, the
pulses first evolve according to the first segments of the dispersion map
for which $\sigma _{j}<0$ (figures 1, 3, 5, and 7). The dynamics then
reverse flow according to the dispersion switches sign as shown in figures
2, 4, 6, and 8, and then finally are again subjects to the dynamics. This
case is in a reflection symmetry of $\beta _{j}=0$ derived from the choice
of dispersion map and the resulting invariance of eqs.(48b), (48c) and (48d)
under the following reversibility transformation $\left( \eta _{j},\beta
_{j},Z\right) \rightarrow \left( \eta _{j},-\beta _{j},-Z\right) .$ The
qualitative desriptions are powerful enough when the exact values of the
fixed points $\left( \eta _{j},\beta _{j}\right) $ (eqs.(63)) are directly
found in the phases-plane dynamics graphics, and have been possibly proved
in this paper.

For the Single \textbf{NLS} investigated before by Kutz, et al., all
explanations about the Hamiltonian dynamics of the dispersion-managed
breathers can be derived from the reduced \textbf{ODEs} of the coupled 
\textbf{NLS }in\textbf{\ section 6}. For instance, the two fixed points and
the eigenvalues which are easily found by reducing eqs.(48) and (59) are as
follows

\begin{equation}
\mathbf{I.}\text{ \ \ \ }\beta _{1}=0,\text{ \ \ \ \ \ }\eta _{1}=0, 
\tag{64a}
\end{equation}

\begin{equation}
\mathbf{II}.\text{ \ \ \ }\beta _{1}=0,\text{ }\eta _{2}=0,\text{\ \ \ \ }%
\eta _{1}=\frac{a_{21}a_{71}}{a_{71}a_{11}}=\eta _{10},  \tag{64b}
\end{equation}
where $\left( b^{2}=\frac{2\kappa }{\sqrt{\frac{\pi }{2}}}\text{and }\rho
=0\right) ,$

\begin{equation}
\eta _{1}=\frac{a_{21}a_{71}}{a_{71}a_{11}}=\frac{A_{1}^{2}}{\sqrt{\pi }%
\kappa \sigma _{1}\left( Z\right) }=\eta _{01},\text{\ \ }  \tag{65}
\end{equation}
and 
\begin{equation}
\lambda _{1,2}=\pm i2\sigma \kappa ^{2}\eta _{01}^{2}  \tag{66}
\end{equation}

\section{\protect\bigskip Conclusion}

In conclusion, we have presented the recent results related to Hamiltonian
Dynamics of Dispersion-Managed Coupled Breathers in Optical Transmission
System. From mathematical viewpoints, the use of a variational method and
its corresponding Hamiltonian formulation of the coupled \textbf{NLS}, the
pulse dynamics in a dispersion-managed optical system can be reduced to four
nonlinear \textbf{ODEs}. By means of the conserved quantity of the
Hamiltonian system, an exact solution of both $\beta _{j}$ and $\eta _{j}$
can also be found in terms of a modified quadrature representations. The
results of the coupled \textbf{NLS} with the perturbation term can also be
reduced\ to both that system which has no the perturbation term and the
single \textbf{NLS} case. From the physical viewpoint, we conclude that a
noise sourced by the amplifier which has a small frequency causes a small
change in the coupled breathers dynamics of the Gaussian pulses. We then get
some phases-plane dynamics of the reduced system in several fixed points. We
have also shown that the linearization and use of the Hamiltonian integral
of the four \textbf{ODEs} describe the dynamics of the phases-plane and
further suggests the possibility of constructing a periodic solution that
depends on the initial amplitude-enhancement factor of the pulse power.
Finally, we then find the proof from the phases-plane dynamics that the
coupled breathers dynamics depent on the certain relationship (eq.(31c)) and
the parameter $A_{j}$, of the chirps and amplitudes.

Thus, based on our results, there are some important open problems related
to the another coupled \textbf{NLS} with a complicated loss and gain terms.
However, the dispersion-managed vector breathers dynamics derived from the
Gaussian pulses as a solution of the nonintegrable vector \textbf{NLS :} 
\[
\begin{array}{c}
i\frac{\partial U_{j}}{\partial Z}+\frac{\sigma _{j}\left( Z\right) }{2}%
\frac{\partial ^{2}U_{j}}{\partial T^{2}}+\left( \sum\limits_{j=1}^{N}\left|
U_{j}\right| ^{2}\right) U_{j}=0, \\ 
\text{where \ \ }\sigma _{1}\left( Z\right) =...=\sigma _{N}\left( Z\right)
=\sigma \left( Z\right)
\end{array}
\]
is in progress. As a result, the advantages for the application of the
Gaussian pulses in optical communication system instead of the initial small
chirps and pulses energy are still open.

\section{ACKNOWLEDGMENTS}

The author (H.I. Elim) would like to thank \textbf{Dr. Chris Daag} for
supporting the research fund coordinated by \textbf{LDF-CIDA/EIUDP} and 
\textbf{Dr.N.N. Akhmediev} for having sent off-prints of the papers related
to the dispersion-managed system and for giving wise advises. The work of
F.P. Zen is funded by Hibah Bersaing, DIKTI, Indonesia.

\section{REFERENCES}

\begin{enumerate}
\item  Kutz, J.N., Holmes, P., Evangelides, S.G., and Gordon, J.P., J. Opt.
Soc. Am. B \textbf{15 }no. 1, 87-95 (1998)

\item  Gabitov, I., Shapiro, E.G., and Turitsyn, S.K., Phys. Rev. E \textbf{%
55}, 3624-3633, (1997)

\item  Okamawari, T., Maruta, A., and Kodama, Y., Opt. Lett. \underline{%
\textbf{23}} no. 9, (1998)

\item  Goldstein, H., \textit{Classical Mechanics} , Addison-Wesley,
Reading, Mass., (1980), Chap.2, 4.

\item  Smith, N.J., Knox, F.M., Doran, N.J., Blow, K.J., and Bennion, I.,
Electron. Lett. \textbf{32}, 54-55, (1996)
\end{enumerate}

\end{document}